\documentclass[twocolumn]{aastex63}

\usepackage{natbib}
\usepackage{hyperref}
\received{XXX}
\revised{XXX}
\accepted{XXX}
\submitjournal{ApJ}

\shorttitle{Spectroscopic study of metal-poor carbon stars}
\shortauthors{Shejeelammal and Goswami}
\graphicspath{{./}{figures}}

\begin{document}

\title{Spectroscopic study of four metal-poor carbon  stars from the Hamburg/ESO Survey: On confirming the low-mass nature of their companions \footnote{Based on data collected using Mercator/HERMES, SUBARU/HDS}}

\correspondingauthor{Aruna Goswami}
\email{aruna@iiap.res.in}

\author[0000-0002-6234-4226]{J. Shejeelammal}
\affiliation{Indian Institute of Astrophysics, Koramangala, Bangalore 560034, India}

\author[0000-0002-8841-8641]{Aruna Goswami}
\affiliation{Indian Institute of Astrophysics, Koramangala, Bangalore 560034, India}

\begin{abstract}
Elemental  abundances of extrinsic carbon stars provide insight into the poorly 
understood origin and evolution of elements in the early Galaxy. In this work, 
we present the results of a detailed spectroscopic analysis of four 
potential carbon star candidates from the Hamburg/ESO Survey (HES)
HE~0457$-$1805, HE~0920$-$0506, HE~1241$-$0337, and HE~1327$-$2116. This analysis 
is based on the high-resolution spectra obtained 
with Mercator/HERMES (R$\sim$86,000) and SUBARU/HDS (R$\sim$50,000). 
Although the abundances of a few elements, such as, Fe, C, 
and O are available  from medium-resolution spectra, we present 
the first ever detailed high-resolution spectroscopic analysis for these objects. The 
object HE~0457$-$1805 and HE~1241$-$0337 are found to be CEMP-s stars, HE~0920$-$0506  a 
CH star, and HE~1327$-$2116  a CEMP-r/s star. The object HE~0457$-$1805 
is a confirmed binary, whereas the binary status of the other objects 
is unknown. The locations of program stars on the absolute carbon abundance  
A(C) vs [Fe/H] diagram point at their binary nature. 
We have examined  various elemental abundance
ratios of the program stars and confirmed the low-mass nature of their former AGB  companions. 
We have shown that the i-process models could successfully reproduce the observed abundance pattern in 
HE~1327$-$2116. The parametric model based analysis performed for HE~0457$-$1805, 
HE~0920$-$0506, and HE~1241$-$0337 based on the FRUITY models  confirmed that the surface chemical composition
of these three objects are influenced by  pollution from  low-mass AGB companions.
\end{abstract}

 
\section{Introduction} \label{section_introduction}
The low- and intermediate-mass stars (0.8 - 10 M$_{\odot}$) are the predominant stellar 
population of our Galaxy \citep{Karakas_2014}.
These stars are the main sites of various nucloesynthesis and important participants of 
chemical evolution of the Universe \citep{Travaglio_2001, Travaglio_2004, Romano_2010, Kobayashi_2011, Kobayashi_2020}. 
As they evolve through different stages of stellar evolution, they enriched the ISM through stellar 
outflows or winds. The pollution from these low- and intermediate- mass stars account for about 90\% 
of the ISM dust and the massive stars account for the rest \citep{Sloan_2008}. 
Majority of the elements heavier than Fe are produced by these stars through slow- (s-) and rapid- (r-) 
neutron-capture processes. The origin and evolution of these heavy elements still remains poorly understood. 
This underscores the need for detailed studies on different classes of stars enhanced in heavy elements, for instance Ba, CH, carbon-enhanced metal-poor (CEMP) stars etc.
The last two
decades had witnessed a significant  increase in the high-resolution 
spectroscopic studies of these groups of  stars.

Ba and CH stars show enhanced abundances of s-process elements. 
The CEMP-stars are metal-poor counterparts ([Fe/H]$<$$-$1) of CH stars (e.g. \citealt{Lucatello_2005, Abate_2016}). 
A fraction of them show enrichment of s- and/or r-elements. They are classified into four sub-classes: 
CEMP-r (enhanced in r-process elements), CEMP-s (enhanced in s-process elements), 
CEMP-r/s (enhanced in both s- and r-process elements), and CEMP-no (little or no enhancement of heavy elements) 
\citep{Beers_2005}. The suggested origin for CEMP-r stars is that they were formed from the ISM pre-enriched by events, such as, 
core-collapse SNe, neutron star mergers, neutron star - black hole mergers 
\citep{Surman_2008, Arcones_2013, Rosswog_2014, Drout_2017, Lippuner_2017}. 
The suggested progenitors for the origin of carbon enhancement of CEMP-no stars, that polluted ISM, 
are faint SNe, spinstar, metal-free massive stars, binary mass-transfer 
from extremely metal-poor AGB stars \citep{Heger_2010, Nomoto_2013, Chiappini_2013, Tominaga_2014}.

CEMP-s stars are metal-poor analog of CH and Ba stars, and binary mass-transfer
from former AGB companion stars have been identified as the source of their observed abundance pattern
(e.g. \citealt{Jorissen_1998, Lucatello_2005, Starkenburg_2014, Jorissen_2016a, Hansen_2016b}). 
However, the exact origin of CEMP-r/s stars is an open question
\citep{Jonsell_2006, Masseron_2010, Koch_2019}. Studies have shown that almost half of the CEMP-s stars 
are CEMP-r/s stars \citep{Sneden_2008, Kappeler_2011, Bisterzo_2011}. Just like the CEMP-s stars, CEMP-r/s stars 
are also found to belong to binary systems \citep{Lucatello_2005, Starkenburg_2014, Hansen_2016b}. 
The scenarios proposed to explain the origin of CEMP-r/s stars include: self-pollution of a star formed 
from r-element enriched ISM, pollution from AGB companion in a binary system formed from r-element enriched ISM, 
binary system polluted from the massive primary in a tertiary system, secondary star polluted from 
the primary exploded as Type 1.5 SN or intermediate neutron-capture (i-) process (\citealt{Jonsell_2006} and references therein, \citealt{Hampel_2016}). The i-process, originally proposed by \cite{Cowan_1977}, produces neutron density 
intermediate between s- and r-process, of the order of 10$^{15-17}$ cm$^{-3}$. There are a number of sites
proposed to host the i-process, such as Rapidly Accreting White Dwarfs \citep{Denissenkov_2017}, 
super-AGB stars of low-metallicity \citep{Doherty_2015, Jones_2016}, metal-poor massive (M $\geq$ 20 M$_{\odot}$)
stars \citep{Banerjee_2018}, extremely low-metallicity (z $\leq$ 10$^{-5}$) low-mass AGB stars \citep{Fujimoto_1990, Hollowell_1990, Lugaro_2009}, low-mass (M $\leq$ 2M$_{\odot}$) very low-metallicity (z $\leq$ 10$^{-4}$) AGB stars 
\citep{Fujimoto_2000, Campbell_2008, Lau_2009, Cristallo_2009, Campbell_2010, Stancliffe_2011}.  etc., though exact site(s) is not confirmed yet. Several studies used i-process in low-mass low-metallicity AGB 
stars to successfully explain the abundances of CEMP-r/s stars \citep{Hampel_2016, Karinkuzhi_2021, Goswami_2021, Shejeelammal_2021a, Shejeelammal_2021b, Purandardas_2021b}.  

\par In this work, we have presented the results of a detailed high-resolution spectroscopic analysis of 
four carbon stars identified from the HES. The structure of the paper is as follows. 
The stellar sample, source of the spectra and data reduction are discussed in Section \ref{section_sample}.
Section \ref{section_RV} provides a discussion on radial velocity. Estimation of stellar atmospheric
parameters, along with the discussion on stellar mass, is provided in Section \ref{section_atmospheric_parameter}. 
In Section \ref{section_abundance_determination}, we have presented the discussion on abundance 
determination, followed by a discussion on abundance uncertainties in Section \ref{section_uncertainty}. 
Classification of program stars is presented in Section \ref{section_classification}. 
A detailed discussion on various abundance ratios of the program stars are provided in 
Section \ref{section_discussion}. A discussion on the origin of the program stars, along 
with the parametric model based analysis, is also given in the 
same section. This section also contains a discussion on individual stars. 
Conclusions are drawn in Section \ref{section_conclusion}.

\section{STELLAR SAMPLE: SELECTION, OBSERVATION/DATA ACQUISITION AND DATA REDUCTION} \label{section_sample}
\begin{figure}
\centering
\includegraphics[width=\columnwidth]{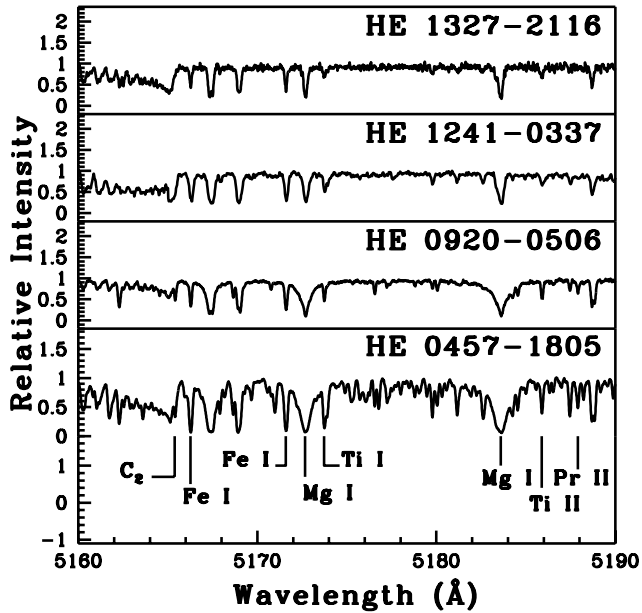}
\caption{Sample spectra of the program stars in the wavelength region 
5160 to 5190 {\bf  {\rm \AA}}. 
}\label{sample spectra}
\end{figure}

All the four objects, HE~0457$-$1805, HE~0920$-$0506, HE1241$-$0337, and HE~1327$-$2116 analysed in this study
are selected from the candidate metal-poor stars identified from the Hamburg/ESO survey (HES) \citep{Christlieb_2003}. 
The stars HE~0457$-$1805 and HE~1241$-$0337 are  listed in the catalog of carbon stars identified from the 
Hamburg/ESO survey (HES) by \cite{Christlieb_2001a}. The object HE~0457$-$1805 is also listed in the list 
of potential CH star candidates by \cite{Goswami_2005}. 
The objects HE~0920$-$0506 and HE~1327$-$2116 are listed in the catalog of bright metal-poor candidates 
from the HES by \cite{Frebel_2006}. The wavelength calibrated, high-resolution (R$\sim$50,000) 
spectrum of HE~1241$-$0337 is taken
from SUBARU/HDS archive (\url{http://jvo.nao.ac.jp/portal/v2/}). The wavelength coverage of
this spectrum is 4100 - 6850 {\rm \AA} with a wavelength gap between 5440 and 5520 {\rm \AA}. 
The high-resolution (R$\sim$86,000) spectra of 
the objects HE~0457$-$1805, HE~0920$-$0506, and HE~1327$-$2116 were obtained using 
the High-Efficiency and high-Resolution Mercator Echelle Spectrograph (HERMES)
attached to the 1.2m Mercator telescope at the Roque de los Muchachos Observatory in
La Palma, Canary Islands, Spain, operated by the Institute of Astronomy of the KU Leuven, Belgium \citep{Raskin_2011}.
 The data were reduced using the HERMES pipeline. The HERMES spectra cover the wavelength range 
 3770 - 9000 {\rm \AA}. Multiple frames of each object were taken on different nights:
15 frames with exposure of 1800 and 2100 sec for HE~0457$-$1805 during the period 2013 - 2016, 
10 frames with exposure ranging from 750 - 1800 sec for HE~0910$-$0506 during 2018 - 2020, and 
3 frames with exposure of 1200 sec for HE~1327$-$2116 on 31/01/2019 and 01/02/2019. 
To maximize the S/N ratio, all these frames were co-added after the Doppler correction. 
The co-added spectra are then  continuum fitted for further analysis, 
using IRAF\footnote{IRAF (Image Reduction and Analysis Facility) 
is distributed by the National Optical Astronomical 
Observatories, which is operated by the Association for Universities 
for Research in Astronomy, Inc., under contract to the National 
Science Foundation} software.
Table \ref{basic data of program stars} provides basic information about the 
program stars, and the sample spectra are shown in Figure \ref{sample spectra}.

 {\footnotesize
\begin{table*}
\caption{\textbf{Basic information of the program stars.}\label{basic data of program stars}}
\resizebox{\textwidth}{!}{
\begin{tabular}{lcccccccccr}
\hline
Star      &RA$(2000)$       &Dec.$(2000)$    &B       &V       &J        &H        &K     &          & S/N       &         \\
          &                 &                &        &        &         &         &      & 4200 \AA &  5500 \AA & 7700 \AA \\
\hline
HE~0457$-$1805  & 04 59 43.56     & $-$18 01 11.99   & 12.372   & 11.014    & 8.937    & 8.421    & 8.186 & 15.82 & 55.48 & 89.05  \\ 
HE~0920$-$0506  & 09 23 05.96     & $-$05 19 32.75   & 11.80    & 10.95     & 10.317   & 9.971    & 9.900 & 14.00 & 51.88 & 49.89 \\ 
HE~1241$-$0337  & 12 44 27.21     & $-$03 54 01.20   & 15.84    & 14.30     & 11.869   & 11.234   & 11.026 & 8.35 & 40.16 & -- \\
HE~1327$-$2116  & 13 30 19.36     & $-$21 32 03.33   & 12.714   & 11.651    & 9.893    & 9.412    & 9.275 & 13.85 & 29.60 & 68.14 \\
\hline

\end{tabular}}
\end{table*}
}

\section{Radial velocity} \label{section_RV}
The radial velocities of the program stars are estimated from the measured shift 
in wavelengths of several spectral lines using the Doppler equation. 
Then, the obtained radial velocity is corrected for heliocentric motion. 
The average value of these heliocentric motion corrected velocities is taken as the radial velocity of the object.
The detailed radial velocity data of the objects analyzed in this study are to be published in a 
summary paper on  orbits of CEMP stars (Jorissen et al., in preparation).
The estimated radial velocities, along with the values from the Gaia DR2 \citep{Gaia_2018} 
and RAVE DR4 \citep{Kordopatis_2013}, are given in Table \ref{atmospheric parameters}. 
The object HE~0457$-$1805 is a confirmed binary with an orbital period of 2724$\pm$23 days \citep{Jorissen_2016a}.
Our radial velocity estimate of this object differ from the Gaia value by $\sim$2 km s$^{-1}$ and 
from the RAVE value by $\sim$9 km s$^{-1}$. 
For the object HE~0920$-$0506, our estimate shows a difference of $\sim$3 km s$^{-1}$ from 
the Gaia value. This may indicate that this star is likely a binary. For the other object, 
HE~1327$-$2116, our estimates show a difference of $\sim$0.7 km s$^{-1}$ from the Gaia and RAVE values. 
The radial velocity estimates for the object HE~1241$-$0337 is not available in Gaia and RAVE.

\section{STELLAR ATMOSPHERIC PARAMETERS} \label{section_atmospheric_parameter}
\begin{figure}
\centering
\includegraphics[width=\columnwidth]{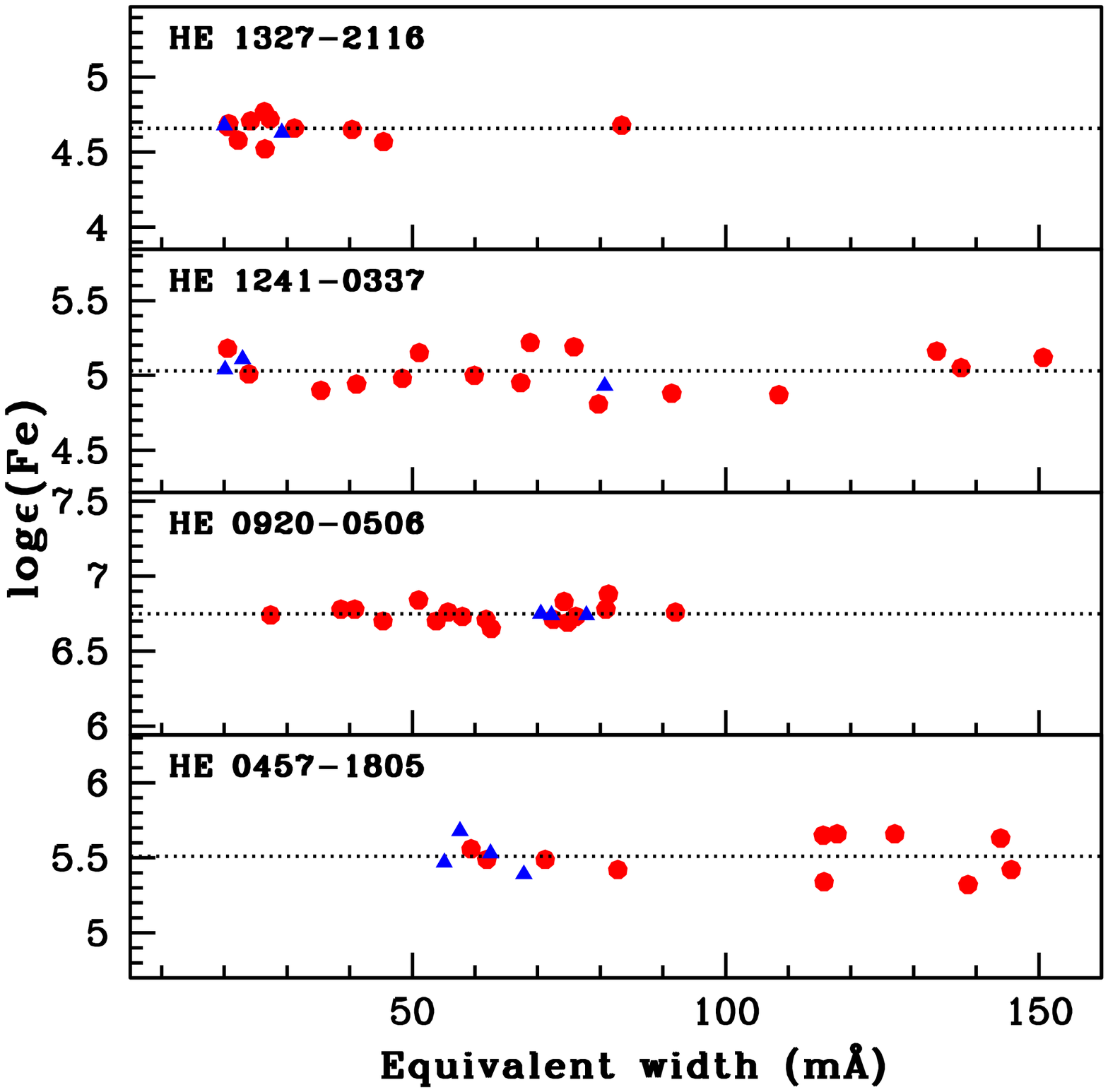}
\includegraphics[width=\columnwidth]{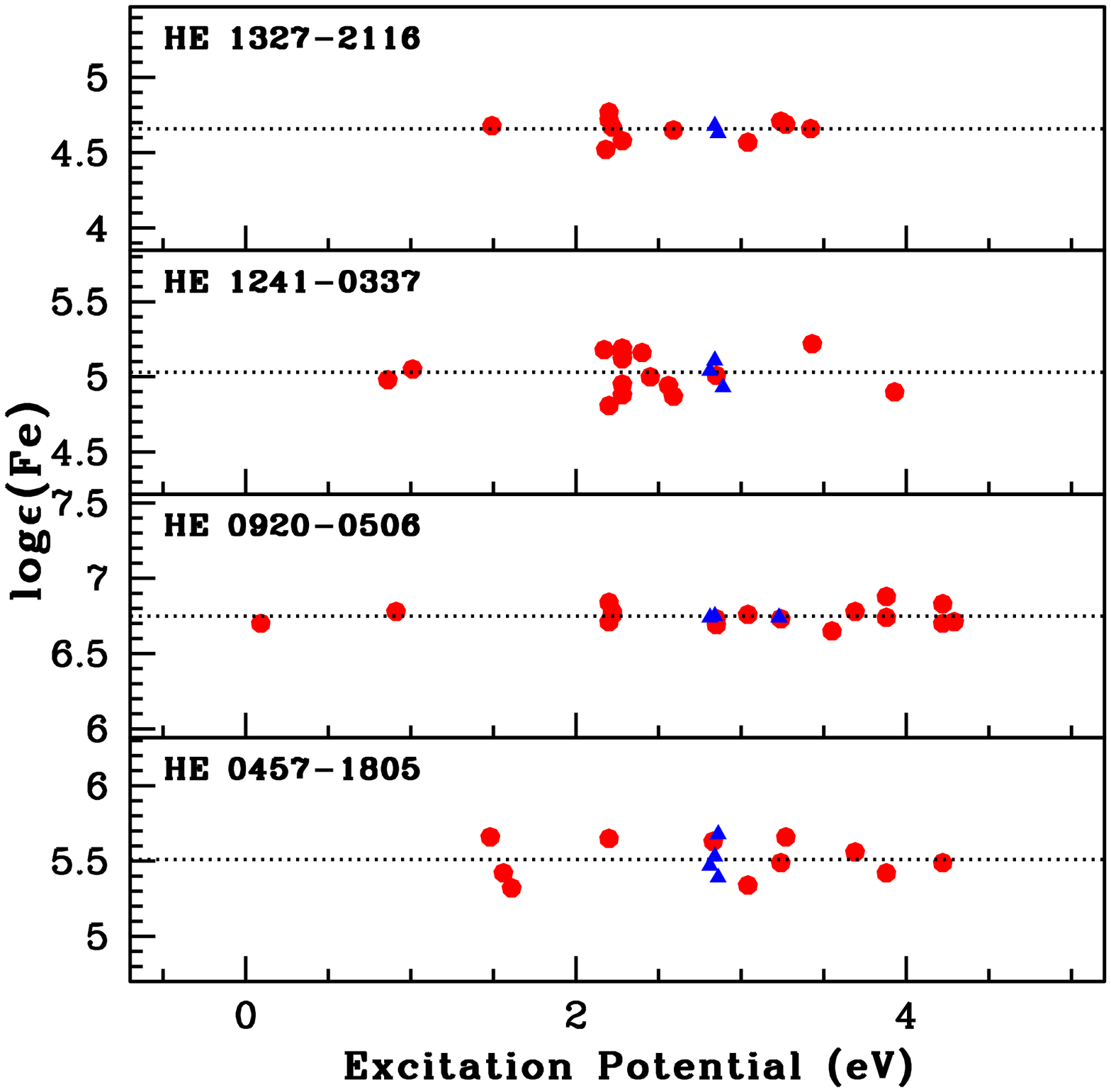}
\caption{Fe abundances of the program stars derived from individual Fe I and Fe II 
lines as function of equivalent width (upper panel),
and excitation potential (lower panel). The adopted value of Fe abundance of each star is shown by
the dotted lines. Solid circles and triangles represent Fe I and Fe II lines respectively. } \label{ep_ew}
\end{figure}

The stellar atmospheric parameters of the program stars 
are derived from the measured equivalent widths of the clean Fe I 
and Fe II lines with excitation potential and equivalent width, 
respectively, in the range of 0 - 6.0 eV and 10 - 180 m{\rm \AA}. 
The radiative transfer code MOOG \citep{Sneden_1973} is used for 
our analysis under the  assumption of  local thermodynamic equilibrium (LTE).
The model atmosphere is selected from the Kurucz grid of model 
atmospheres with no convective overshooting (\citealt{Castelli_2003}, \url{http://kurucz.harvard.edu/grids.html}).
To select the initial model atmosphere to start with, we have used the 
photometric temperature estimates calculated from the calibration equations 
given by \cite{Alonso_1999, Alonso_2001} and a guess of typical log g value for giants.
Using the usual excitation balance and ionization balance method, the final model atmosphere is 
obtained through an iterative process from the initial one. The detailed procedure 
is described in our earlier papers \cite{Shejeelammal_2020, Shejeelammal_2021a, Shejeelammal_2021b}. 
Derived Fe abundances of the program stars are shown in Figure \ref{ep_ew},  
and their derived atmospheric parameters are given in Table \ref{atmospheric parameters}. 
The literature values of these atmospheric parameters are also provided in the same table. 
The comparison of our estimates of stellar atmospheric parameters 
with the literature values is discussed in Section \ref{section_individual_stars}.

{\footnotesize
\begin{table*}
\caption{Derived atmospheric parameters and radial velocity of the program stars.} \label{atmospheric parameters}
\resizebox{\textwidth}{!}{
\begin{tabular}{lccccccccc}
\hline
Star                &T$\rm_{eff}$  & log g      &$\zeta$         & [Fe I/H]          &[Fe II/H]          & V$_{r}$             &  V$_{r}$  & V$_{r}$  & Reference \\
                    &    (K)       & cgs        &(km s$^{-1}$)   &                   &                   & (km s$^{-1}$)       & 
                    (km s$^{-1}$)  & (km s$^{-1}$) &  \\
                    & $\pm$100     & $\pm$0.2   & $\pm$0.2       &                   &                   & (This work)         & (Gaia)      & RAVE     &    \\  
\hline
HE~0457$-$1805      & 4435      & 0.70  & 1.97  & $-$1.98$\pm$0.13  & $-$1.98$\pm$0.12  & +62.83$\pm$0.02   & +60.80$\pm$0.35  &     72.00$\pm$2.2 & 1  \\
                    & 4484      & 0.77  & --    & $-$1.46           & --                & --                & --               & -- 
                    & 2  \\
HE~0920$-$0506      & 5380      & 2.65  & 0.69  & $-$0.75$\pm$0.06  & $-$0.75$\pm$0.01  & +49.60$\pm$0.03   & +52.44$\pm$1.27  & --              & 1  \\
                    & --      & --  & --    & $-$1.01           & --                & --                & --               & --      & 3 \\
                    & 5291      & 2.99  & --    & $-$1.39           & --                & --                & --               & --      & 4 \\
HE~1241$-$0337      &   4240    & 1.18  & 2.82  & $-$2.47$\pm$0.13  & $-$2.47$\pm$0.09  & +179.69$\pm$2.18  & --   & -- &  1  \\                
HE~1327$-$2116      & 4835      & 1.50  & 3.45  & $-$2.84$\pm$0.07  & $-$2.84$\pm$0.05  & +176.77$\pm$0.06  & +177.43$\pm$1.46 & 177.4$\pm$1.5 & 1  \\
                    & --       & --  & --    & $-$2.93           & --                & --                & --               & --      & 3 \\
                   & 4868      & 0.55  & --    & $-$3.48           & --                & --                & --               & --      & 4 \\
\hline
\end{tabular}
} 

References: 1. Our work, 2. \cite{Kennedy_2011}, 3. \cite{Frebel_2006}, 4. \cite{Beers_2017} \\
\end{table*}
}

\begin{figure}
\centering
\includegraphics[width=\columnwidth]{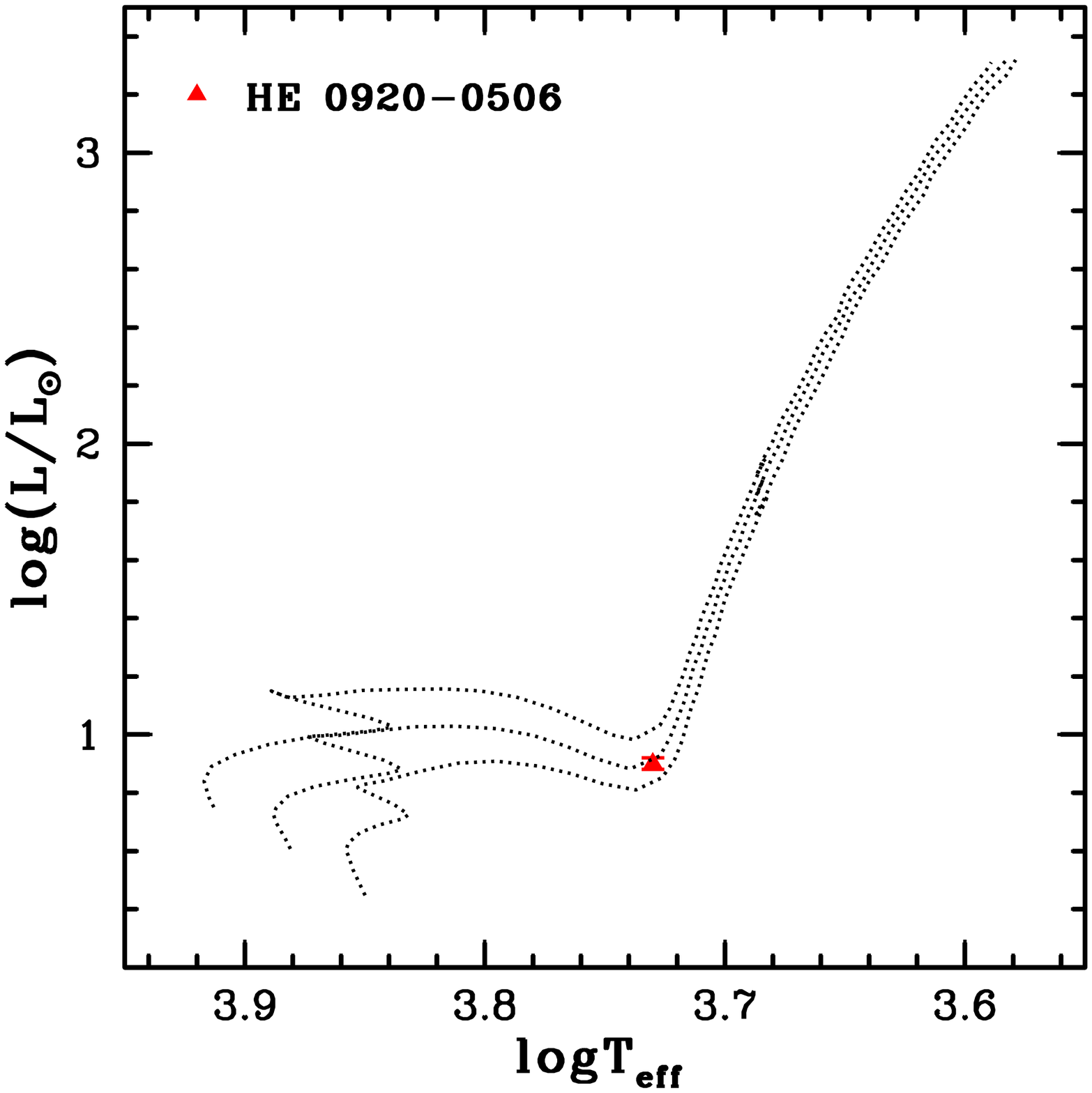}
\includegraphics[width=\columnwidth]{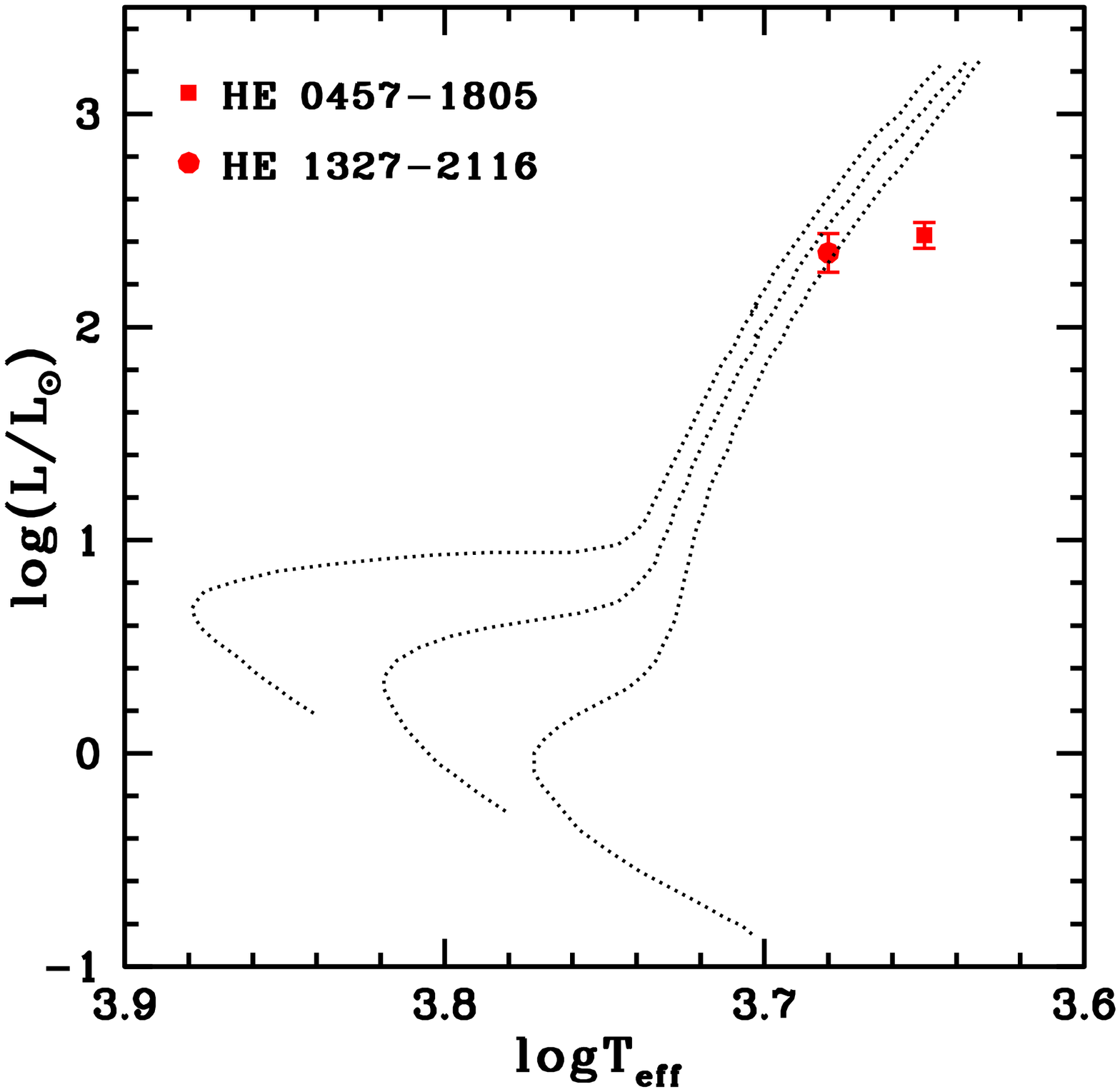}
\caption{The evolutionary tracks for 1.0, 1.1, and 1.4 M$_{\odot}$ for z = 0.004 (upper panel)  
and for 0.6, 0.8, and  1.0 M$_{\odot}$ for z = 0.0004 (lower panel)
are shown from bottom to top.} \label{tracks}
\end{figure}

The surface gravity is also calculated from the parallax method using \\
log (g/g$_{\odot}$)= log (M/M$_{\odot}$) + 4log (T$\rm_{eff}$/T$\rm_{eff\odot}$) - log (L/L$_{\odot}$)\\ 
The adopted solar values are log g$_{\odot}$ = 4.44, T$\rm_{eff\odot}$ = 5770 K, and M$\rm_{bol\odot}$ = 4.74 mag.
The masses of the program stars are found from their positions on the H-R diagram (log T$\rm_{eff}$ - log (L/L$_{\odot}$) diagram)
generated using the evolutionary tracks of \cite{Girardi_2000}. 
This procedure is discussed in detail in \cite{Shejeelammal_2020}. 
For the objects HE~0457$-$1805 and HE~1327$-$2116, z = 0.0004 ([Fe/H]$\sim$$-$1.7) tracks were used and 
z = 0.004 ([Fe/H]$\sim$$-$0.7) tracks for HE~0920$-$0506. The positions of the program stars on the H-R diagram are 
shown in Figure \ref{tracks}. 
We could find only an upper limit to the 
mass of the star HE~0457$-$1805. We could not estimate the mass of the star HE~1241$-$0337 
since its parallax value is not available. The estimated mass and log\, g values of the program stars 
are given in Table \ref{mass_CEMP_stars}. For the star HE~0920$-$0506, the spectroscopic log\, g
is $\sim$0.8 dex lower than that estimated from the parallax method. 
In some carbon stars, such an inconsistency between the spectroscopic atmospheric parameters and
those derived from the evolutionary tracks could arise as their evolutionary tracks shift
towards lower temperatures \citep{Marigo_2002, Jorissen_2016b}. 
The spectroscopic log\, g values have been used for our analysis. 

{\footnotesize
\begin{table*}
\caption{Mass and log\,{g} estimated from parallax method} \label{mass_CEMP_stars}
\begin{tabular}{lcccccc}
\hline                       
 Star name         & Parallax             & M$\rm_{bol}$           & log(L/L$_{\odot}$)  & Mass(M$_{\odot}$) & log g          & log g (spectroscopic)   \\
                   & (mas)                &                     &                     &                   & (cgs)          & (cgs)                    \\
\hline
HE~0457$-$1805     & 0.467$\pm$0.0309     & $-$1.341$\pm$0.14   & 2.43$\pm$0.05       & $<$0.60           & --             & 0.70  \\ 
HE~0920$-$0506     & 2.3764$\pm$0.0653    & 2.488$\pm$0.06      & 0.90$\pm$0.02       & 1.30$\pm$0.005    & 3.50$\pm$0.03  & 2.65  \\ 
HE~1327$-$2116     & 0.3428$\pm$0.0358    & $-$1.130$\pm$0.228  & 2.35$\pm$0.09       & 0.70$\pm$0.10     & 1.63$\pm$0.03  & 1.50  \\
\hline
\end{tabular}
\end{table*}
}

\section{Abundance determination} \label{section_abundance_determination}
The elemental abundances were estimated from two methods: (i) using the equivalent 
widths of the spectral lines and (ii) by comparison of observed spectra and synthetic 
spectra generated with MOOG (spectral synthesis calculation), using Kurucz model 
atmospheres (\url{http://kurucz.harvard.edu/grids.html}). The spectral synthesis calculation 
is performed to derive the abundances of elements with hyper-fine splitting (HFS) as well as 
for molecular bands. The hyper-fine components of each atomic line, whenever available, 
are considered for the abundance estimation. In the case of abundance determination using 
the equivalent width method, the spectral line identification is performed by comparing the 
stellar spectra with the Doppler corrected spectrum of Arcturus. 
The line parameters are taken from linemake \footnote{linemake contains 
laboratory atomic data (transition probabilities, hyperfine and isotopic substructures) 
published by the Wisconsin Atomic Physics and the Old Dominion Molecular Physics groups. 
These lists and accompanying line list assembly software have been developed by 
C. Sneden and are curated by V. Placco at \url{https://github.com/vmplacco/linemake}.} 
atomic and molecular line database \citep{Placco_2021}.  

Although the abundance 
estimation is performed under the assumption of LTE, we have applied the non-LTE (NLTE) 
corrections whenever available. The solar abundance values are adopted from \cite{Asplund_2009}.
The details of the abundances of each elements and the details of NLTE corrections as well as 
HFS are discussed in this section. The estimated elemental abundances in the program stars 
are given in Tables \ref{abundance_table1} and \ref{abundance_table2}. The atomic lines used to derive the abundances of 
each elemental species are given in Tables \ref{linelist1} - \ref{linelist3}.

{\footnotesize
\begin{table*}
\caption{Elemental abundances in HE~0457$-$1805 and HE~0920$-$0506} \label{abundance_table1}
\resizebox{\textwidth}{!}
{\begin{tabular}{lcccccccccc}
\hline
       &    &                                &                    & HE~0457$-$1805 &       & &                    & HE~0920$-$0506 &  & \\ 
\hline
       & Z  & solar log$\epsilon^{\ast}$   & log$\epsilon$      & [X/H]    & [X/Fe]  & N   & log$\epsilon$      & [X/H]    & [X/Fe] & N   \\ 
\hline 
C$^{\dagger}$ (C$_{2}$ band 5165 {\rm \AA})	&	6	&	8.43	&	8.32	&	$-$0.11	&	1.87 & --	&	8.20	&	$-$0.23	&	0.52	& -- \\
C$^{\dagger}$ (C$_{2}$ band 5635 {\rm \AA})	&	6	&	8.43	&	8.17	&	$-$0.26	&	1.72 & --	&	8.30	&	$-$0.13	&	0.62 & --	\\
$^{12}$C/$^{13}$C$^{\dagger}$	&	--	&	--	&	--	&	--	&	23$\pm$4 & --	&	--	&	--	&	--	& --	\\
N$^{\dagger}$ 	&	7	&	7.83	&	7.52$\pm$0.07	&	$-$0.31	&	1.67 & 3	&	7.53$\pm$0.20	&	$-$0.30	&	0.45	& 3 \\
O$^{\dagger}$	&	8	&	8.69	&	--	&	--	&	-- & --	&	7.80	&	$-$0.89	&	$-$0.14	& 1	\\
Na I	&	11	&	6.24	&	6.46$\pm$0.06	&	0.22	&	2.2	& 3 &	5.94$\pm$0.09	&	$-$0.30	&	0.45 & 3	\\
Mg I	&	12	&	7.6	&	6.94$\pm$0.14	&	$-$0.66	&	1.32 & 2	&	7.50$\pm$0.03	&	$-$0.10	&	0.65 & 3	\\
Si I	&	14	&	7.51	&	5.98$\pm$0.17	&	$-$1.53	&	0.45 & 2	&	7.14$\pm$0.17	&	$-$0.37	&	0.38 & 2	\\
Ca I	&	20	&	6.34	&	5.21$\pm$0.12	&	$-$1.13	&	0.85  & 7	&	5.98$\pm$0.11	&	$-$0.36	&	0.39 & 16	\\
Sc II$^{\dagger}$	&	21	&	3.15	&	1.85	&	$-$1.30	&	0.68 & 1	&	2.14	&	$-$1.09	&	$-$0.26	& 2	\\
Ti I	&	22	&	4.95	&	3.83$\pm$0.17	&	$-$1.12	&	0.86 & 6	&	4.40$\pm$0.08	&	$-$0.55	&	0.2	& 8	\\
Ti II	&	22	&	4.95	&	3.40$\pm$0.14	&	$-$1.55	&	0.43 & 5	&	4.34$\pm$0.08	&	$-$0.61	&	0.14	& 8	\\
V I$^{\dagger}$	&	23	&	3.93	&	2.73	&	$-$1.20	&	0.78 & 2	&	2.87	&	$-$1.06	&	$-$0.31	& 1	\\
Cr I	&	24	&	5.64	&	4.72$\pm$0.10	&	$-$0.92	&	1.06 & 4	&	5.36$\pm$0.09	&	$-$0.28	&	0.47 & 10		\\
Cr II	&	24	&	5.64	&	4.58$\pm$0.04	&	$-$1.06	&	0.92 & 2	&	5.37$\pm$0.07	&	$-$0.27	&	0.48 & 3		\\
Mn I$^{\dagger}$	&	25	&	5.43	&	4.43	&	$-$1.00	&	0.98 & 2	&	4.44$\pm$0.01	&	$-$0.99	&	$-$0.24	& 2	\\
Fe I	&	26	&	7.5	&	5.52$\pm$0.13	&	$-$1.98	&	-- & 11	&	6.75$\pm$0.06	&	$-$0.75	&	-- & 17		\\
Fe II	&	26	&	7.5	&	5.52$\pm$0.12	&	$-$1.98	&	-- & 4	&	6.75$\pm$0.01	&	$-$0.75	&	-- & 3		\\
Co I$^{\dagger}$	&	27	&	4.99	&	3.58	&	$-$1.41	&	0.57 & 1	&	4.19	&	$-$0.80	&	$-$0.05 & 1		\\
Ni I	&	28	&	6.22	&	5.30$\pm$0.08	&	$-$0.92	&	1.06 & 7	&	5.83$\pm$0.10	&	$-$0.39	&	0.36 & 10		\\
Cu I$^{\dagger}$	&	29	&	4.19	&	2.21	&	$-$1.98	&	0 & 1	&	3.19	&	$-$1.00	&	$-$0.25	& 1	\\
Zn I	&	30	&	4.56	&	3.10	&	$-$1.50	&	0.48 & 1	&	4.34$\pm$0.12	&	$-$0.22	&	0.53 & 2		\\
Rb I$^{\dagger}$ 	&	37	&	2.52	&	1.95	&	$-$0.57	&	1.41 & 1	&	1.70	&	$-$0.82	&	$-$0.07	& 1	\\
Sr I$_{NLTE}$$^{\dagger}$	&	38	&	2.87	&	2.54	&	$-$0.33	&	1.65 & 1	&	3.57	&	0.7	&	1.45 & 1	\\
Y I$^{\dagger}$	&	39	&	2.21	&	3.01	&	0.8	&	2.78 & 1	&	2.32	&	0.11	&	0.86 & 1		\\
Y II	&	39	&	2.21	&	2.18$\pm$0.20	&	$-$0.03	&	1.95 & 4	&	2.67$\pm$0.11	&	0.46	&	1.21 & 6		\\
Zr I$^{\dagger}$	&	40	&	2.58	&	3.03	&	0.45	&	2.43 & 1	&	3.05	&	0.47	&	1.22 & 1		\\
Zr II$^{\dagger}$	&	40	&	2.58	&	2.63	&	0.05	&	2.03 & 1	&	2.38	&	$-$0.20	&	0.55 & 1		\\
Ba II$_{LTE}$$^{\dagger}$	&	56	&	2.18	&	--	&	--	&   --	& -- &	2.83	&	0.65	&	1.4	& 1 	\\
Ba II$_{NLTE}$$^{\dagger}$	&	56	&	2.18	&	2.73	&	0.55	&	2.53 & 1	&	2.83	&	0.65	&	1.4	& 1	\\
La II$^{\dagger}$	&	57	&	1.1	&	1.39$\pm$0.02	&	0.29	&	2.27 & 3	&	1.60	&	0.5	&	1.25 & 2			\\
Ce II	&	58	&	1.58	&	1.95$\pm$0.15	&	0.37	&	2.35 & 6	&	1.99$\pm$0.14	&	0.41	&	1.16 & 6		\\
Pr II	&	59	&	0.72	&	1.30$\pm$0.11	&	0.58	&	2.56 & 6	&	1.15$\pm$0.03	&	0.43	&	1.18 & 2		\\
Nd II	&	60	&	1.42	&	1.81$\pm$0.11	&	0.39	&	2.37 & 9	&	1.74$\pm$0.14	&	0.32	&	1.07 & 4		\\
Sm II	&	62	&	2.41	&	1.39$\pm$0.11	&	0.43	&	2.41 & 7	&	1.57$\pm$0.18	&	0.61	&	1.36 & 4		\\
Eu II$_{LTE}$$^{\dagger}$	&	63	&	0.52	&	$-$0.20	&	$-$0.72	&	1.26 & 1	&	$-$0.39	&	$-$0.91	&	$-$0.16		 & 1\\
Eu II$_{NLTE}$$^{\dagger}$	&	63	&	0.52	&	--	&	--	&	--	& -- &	$-$0.43	&	$-$0.95	&	$-$0.20	& 1	\\
\hline
\end{tabular}}

$\ast$  \cite{Asplund_2009}. $^{\dagger}$ indicates that the abundances are derived from spectral synthesis method. 
N is the number of lines used to derive the abundance. 
NLTE refers to the abundance derived from the lines affected by NLTE, after the corrections being applied. 
\end{table*}
}

{\footnotesize
\begin{table*}
\caption{Elemental abundances in HE~1241$-$0337 and HE~1327$-$2116} \label{abundance_table2}
\resizebox{\textwidth}{!}
{\begin{tabular}{lcccccccccc}
\hline
       &    &                                &        & HE~1241$-$0337 &              &   &   & HE~1327$-$2116       &       & \\ 
\hline
       & Z  & solar log$\epsilon^{\ast}$   & log$\epsilon$      & [X/H]    & [X/Fe]  & N   & log$\epsilon$      & [X/H]    & [X/Fe]   & N  \\ 
\hline 
C$^{\dagger}$ (C$_{2}$ band 5165 {\rm \AA})	&	6	&	8.43	&	--	&	--	&	--	& -- &	8.05	&	$-$0.38	&	2.46 & --		\\
C$^{\dagger}$ (C$_{2}$ band 5635 {\rm \AA})	&	6	&	8.43	&	8.53	&	0.10	&	2.57 & --	&	8.05	&	$-$0.38	&	2.46 & --\\
$^{12}$C/$^{13}$C$^{\dagger}$	&	--	&	--	&	--	&	--	& -- &	--	&	--	&	--	&	7$\pm$3	& --	\\
N$^{\dagger}$ 	&	7	&	7.83	&	6.38	&	$-$1.45	&	1.02 & --	&	7.50$\pm$0.10	&	$-$0.33	&	2.51 & 3		\\
Na I	&	11	&	6.24	&	5.35$\pm$0.18	&	$-$0.89	&	1.58 & 2	&	4.28	&	$-$1.96	&	0.88 & 1		\\
Mg I	&	12	&	7.6	    &	5.17$\pm$0.13	&	$-$2.43	&	0.04 & 2	&	5.22	&	$-$2.38	&	0.46 & 1		\\
Si I	&	14	&	7.51	&	5.82$\pm$0.10	&	$-$1.69	&	0.78 & 2	&	5.45	&	$-$2.06	&	0.78 & 1		\\
Ca I	&	20	&	6.34	&	4.15$\pm$0.15	&	$-$2.19	&	0.28 & 6	&	4.13$\pm$0.06	&	$-$2.21	&	0.63 & 5		\\
Sc II$^{\dagger}$	&	21	&	3.15	&	1.00$\pm$0.13	&	$-$2.15	&	0.32 & 2	&	0.03$\pm$0.01 &	$-$3.12	&	$-$0.27	& 2	\\
Ti I	&	22	&	4.95	&	2.77$\pm$0.12	&	$-$2.18	&	0.29 & 3	&	2.77$\pm$0.02	&	$-$2.18	&	0.66 & 2		\\
Ti II	&	22	&	4.95	&	2.53$\pm$0.19	&	$-$2.42	&	0.05 & 4	&	2.59$\pm$0.08	&	$-$2.36	&	0.48 & 4		\\
Cr I	&	24	&	5.64	&	--	&	--	&	--	& -- &	3.03$\pm$0.01	&	$-$2.61	&	0.23 & 2		\\
Cr II	&	24	&	5.64	&	3.55$\pm$0.03	&	$-$2.09	&	0.38 & 2	&	--	&	--	&	-- & --		\\
Mn I$^{\dagger}$	&	25	&	5.43	&	3.55	&	$-$1.88	&	0.59 & 1	&	3.61$\pm$0.03	&	$-$1.81	&	1.03 & 2		\\
Fe I	&	26	&	7.5	    &	5.03$\pm$0.13	&	$-$2.47	&	-	& 16 &	4.66$\pm$0.07	&	$-$2.84	&	- & 11		\\
Fe II	&	26	&	7.5	    &	5.03$\pm$0.09	&	$-$2.47	&	-	& 3 &	4.66$\pm$0.04	&	$-$2.84	&	- & 2		\\
Co I	&	27	&	4.99	&	--	&	--	&	--	& -- &	1.81$\pm$0.07	&	$-$3.18	&	$-$0.34 & 2		\\
Ni I	&	28	&	6.22	&	3.64$\pm$0.15	&	$-$2.58	&	$-$0.11	& 4 &	4.69$\pm$0.06	&	$-$1.53	&	1.31 & 3		\\
Zn I	&	30	&	4.56	&	1.81	&	$-$2.75	&	$-$0.28 & 1	&	2.16	&	$-$2.40	&	0.44 & 1		\\
Sr I$_{NLTE}$$^{\dagger}$	&	38	&	2.87	&	1.54	&	$-$1.33	&	1.14 & 1	&	--	&	--	&	--	& --	\\
Y II	&	39	&	2.21	&	0.86$\pm$0.11	&	$-$1.35	&	1.12 & 5	&	0.06$\pm$0.02	&	$-$2.15	&	0.69 & 2		\\
Zr I$^{\dagger}$	&	40	&	2.58	&	1.44	&	$-$1.14	&	1.33 & 1	&	1.81	&	$-$0.77	&	2.07 & 1	\\
Zr II	&	40	&	2.58	&	1.54$\pm$0.04	&	$-$1.04	&	1.43 & 4	&	0.89$^{\dagger}$	&	$-$1.69	&	1.15 & 1		\\
Ba II$_{NLTE}$$^{\dagger}$	&	56	&	2.18	&	0.75	&	$-$1.43	&	1.04 & 1 &	1.08	&	$-$1.10	&	1.74	& 1	\\
La II$^{\dagger}$	&	57	&	1.1	&	$-$0.35	& $-$1.45	&	1.02 & 1	&	0	&	$-$1.10	&	1.74 & 1		\\
Ce II	&	58	&	1.58	&	0.27$\pm$0.19	&	$-$1.31	&	1.16 & 3	&	0.49$\pm$0.10	&	$-$1.09	&	1.75 & 5		\\
Pr II	&	59	&	0.72	&	$-$0.43$\pm$0.12	&	$-$1.15	&	1.32 & 3	&	$-$0.23	&	$-$0.95	&	1.89 & 1		\\
Nd II	&	60	&	1.42	&   $-$0.03$\pm$0.12	&	$-$1.45	&	1.02 & 4	&	0.36$\pm$0.08	&	$-$1.06	&	1.78 & 8		\\
Sm II	&	62	&	2.41	&	$-$0.52$\pm$0.01	&	$-$1.48	&	0.99 & 2	&	$-$0.07$\pm$0.15	&	$-$1.03	&	1.81 & 4	\\
Eu II$_{LTE}$$^{\dagger}$	&	63	&	0.52	&	$-$1.43$\pm$0.15	&	$-$1.95	&	0.52 & 2	&	$-$1.16	&	$-$1.68	&	1.16	& 1	\\
\hline
\end{tabular}}

$\ast$  \cite{Asplund_2009}. $^{\dagger}$ indicates that the abundances are derived from spectral synthesis method. 
N is the number of lines used to derive the abundance. 
NLTE refers to the abundance derived from the lines affected by NLTE, after the corrections being applied. 
\end{table*}
}

\subsection{Light elements: C, N, O, $^{12}$C/$^{13}$C, Na, $\alpha$-, and Fe-peak elements}
The oxygen abundance could be derived only in HE~0920$-$0506 where
we have used the spectral synthesis calculation of [O I] line at 6300.304 {\rm \AA}.
Oxygen is found to be slightly under abundant in this star with a value [O/Fe]$\sim$$-$0.14. 
We could not detect good [O I] 6300.304  {\rm \AA} line in other stars. 
The [O I] line at 6363.776 {\rm \AA} and oxygen triplet at 7770  {\rm \AA}
were not usable for abundance determination in any of the program stars. 

\par
\begin{figure}
\centering
\includegraphics[width=\columnwidth]{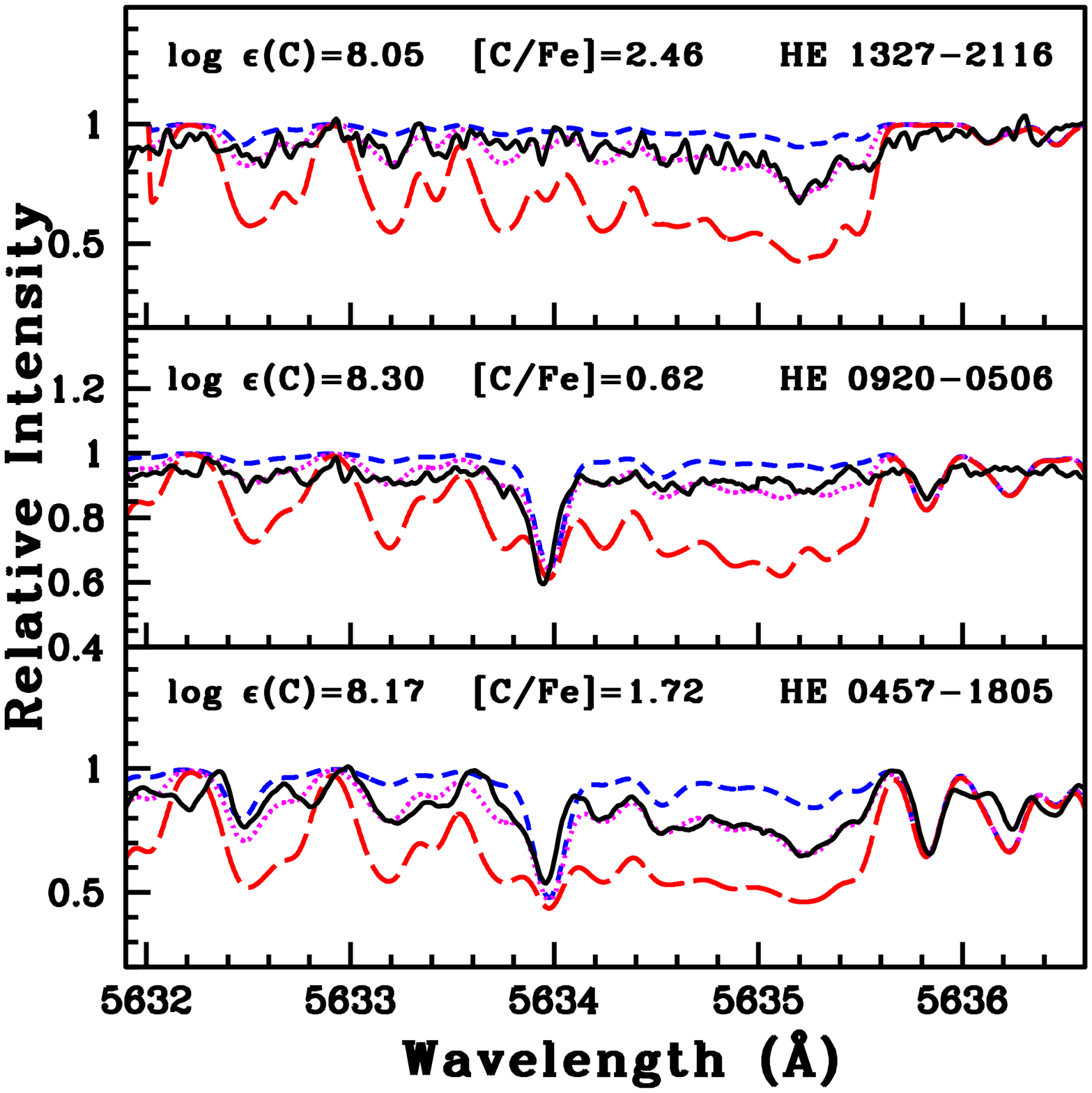}
\includegraphics[width=\columnwidth]{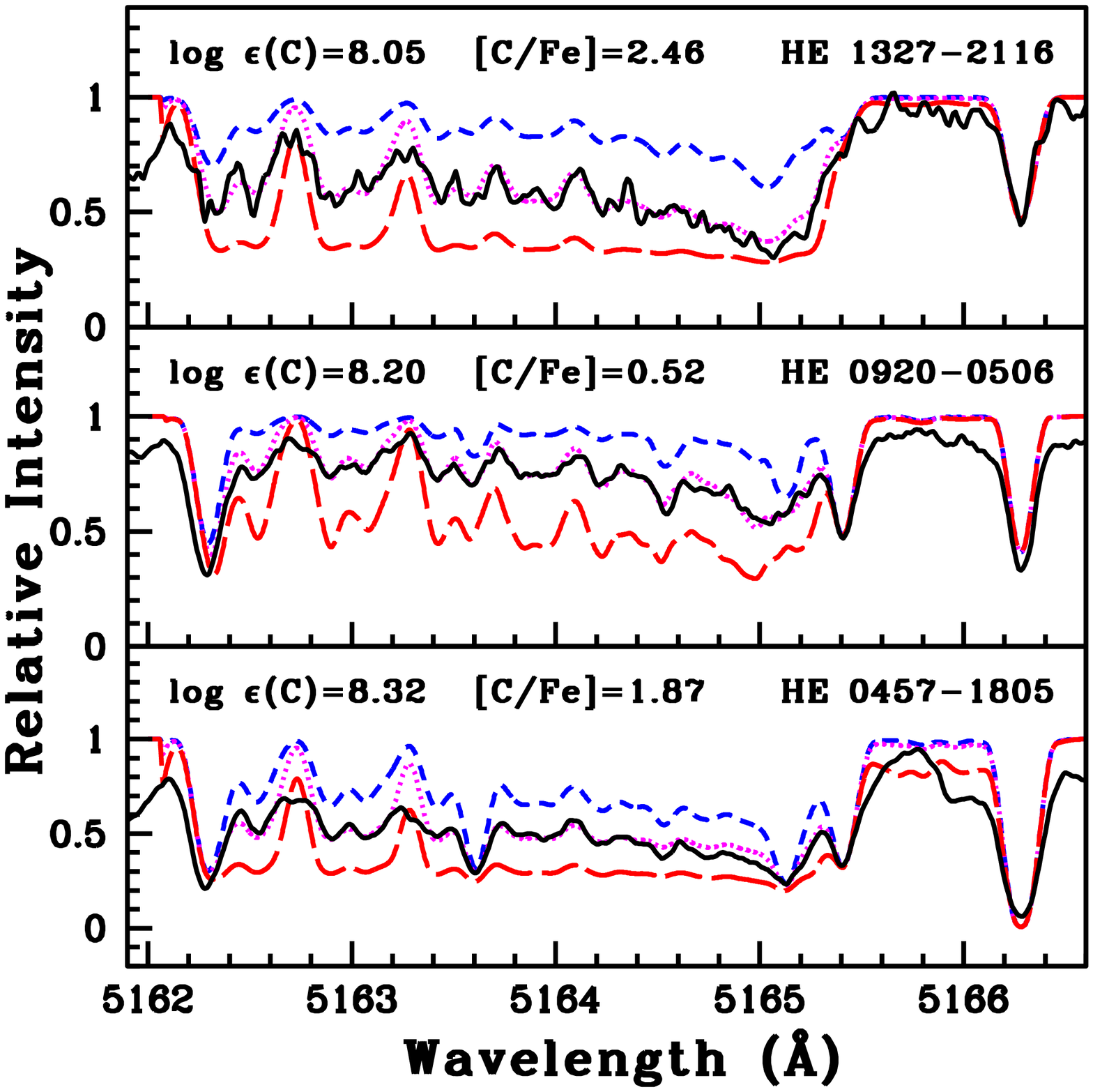}
\caption{ Spectral synthesis fits of C$_{2}$ band around 5165 {\rm \AA} (lower panel) 
and 5635 {\rm \AA} (upper panel). 
Dotted and solid lines respectively represent synthesized and observed spectra. 
short-dashed and long-dashed lines are the synthetic spectra 
for $\Delta$ [C/Fe] = $-$0.3 and +0.3 respectively.} \label{carbon}
\end{figure}

 We could determine the carbon abundance in all the four program stars. 
The carbon abundance is derived from the spectral synthesis calculation of 
C$_{2}$ bands around 5165 and 5635 {\rm \AA} except for HE~1241$-$0337 where the C$_{2}$ 5165 {\rm \AA}
band is noisy. The carbon abundances estimated from 
these two bands are consistent within 0.15 dex, and the final carbon abundance is taken
to be the average of these two abundance values. While the star HE~0920$-$0506 shows a moderate 
enhancement of carbon with [C/Fe]$\sim$0.57, the other three stars are enhanced in carbon with 
[C/Fe]$>$1.70. The spectral synthesis fits of the two carbon bands in the program stars
are shown in Figure \ref{carbon}. 

\par We could estimate nitrogen in all the program stars. 
In HE~1241$-$0337, the nitrogen abundance is estimated from the spectrum synthesis of 
$^{12}$CN band at 4215 {\rm \AA}. 
In the other three program stars, nitrogen abundance is derived from the spectral 
synthesis calculation of $^{12}$CN lines around 8000 {\rm \AA}. 
Among the program stars, the object HE~1327$-$2116 shows the highest enhancement of 
nitrogen with [N/Fe]$\sim$2.51.  

\par The carbon isotopic ratio, $^{12}$C/$^{13}$C, is derived using the spectral 
synthesis calculation of $^{12}$CN and $^{13}$CN lines around 8000 {\rm \AA}. 
We could not estimate the value of $^{12}$C/$^{13}$C ratio in HE~0920$-$0506 
as the $^{13}$CN lines were not good, and in HE~1241$-$0337 as this region is not present in its spectrum. 
The values obtained for this ratio in the
stars HE~0457$-$1805 and HE~1327$-$2116 are 23$\pm$4 and 7$\pm$3 respectively.

\par The abundances of elements Na, Mg, Si, Ca, Ti, Cr, Ni, and Zn are 
estimated from the measured equivalent widths of spectral lines 
listed in Table \ref{linelist2}. We could estimate the abundances of
all these elements in all the program stars. 
The object HE~0457$-$1805 shows the largest enhancement of Na with 
[Na/Fe]$\sim$2.20. While HE~1241$-$0337 and HE~1327$-$2116 show an enhancement of 
[Na/Fe]$\sim$1.58 and 0.88 respectively, 
Na is moderately enhanced in HE~0920$-$0506 with [Na/Fe]$\sim$0.45. 
In very metal-poor stars, Na suffers large uncertainties due to NLTE corrections or
3D hydrodynamical model atmospheres \citep{Bisterzo_2011,Andrievsky_2007}. 
The NLTE effect may reduce the Na abundance by up to $\sim$0.7 dex \citep{Andrievsky_2007}.
However, the Na I 5682.633, 5688.205, 6154.226 and 6160.747 {\rm \AA} lines have negligible 
NLTE effect \citep{Takeda_2003, Lind_2011}.
We have used these weak lines to derive the Na abundance. 
The Mg abundance is derived mainly from Mg I 4702.991, 5528.405, 5711.088 {\rm \AA} 
lines. The object HE~0457$-$1805 is enhanced in Mg as well with [Mg/Fe]$\sim$1.32. 
The object HE~1241$-$0337 shows a near-solar value, whereas the other two objects 
show moderate enhancement of Mg. 

\par We have used the spectral synthesis calculation to derive the abundances 
of elements Sc, V, Mn, Co and Cu by incorporating their hyper-fine
components, whenever available. 
The Sc II 6245.637 {\rm \AA} line is used to derive Sc abundance in HE~0457$-$1805, 
Sc II 6245.637, 6604.601 {\rm \AA} lines were used HE~0920$-$0506. 
The Sc II lines at 4320.732, 4415.556 {\rm \AA} 
were used in HE~1241$-$0337. 
In the case of HE~1327$-$2116, Sc II 4374.457, 5031.021 {\rm \AA} lines were used. 
In HE~0920$-$0506 and HE~1327$-$2116, scandium is slightly under abundant with 
[Sc/Fe]$\sim-$0.25, while it is moderately enhanced in other two stars.
Vanadium abundance is derived from the spectral synthesis calculation of 
V I 4864.731, 5727.048 {\rm \AA} lines. We  could not estimate vanadium abundance 
in HE~1241$-$0337 and HE~1327$-$2116 as there is no clean lines available. 
The objects HE~0457$-$1805 and HE~0920$-$0506 show 
[V/Fe] value 0.78 and $-$0.31  respectively. Manganese abundance is derived from the 
spectral synthesis calculation of Mn I 6013.513, 6021.89 {\rm \AA} lines in the stars 
HE~0457$-$1805 and HE~0920$-$0506. The Mn I 5516.743 {\rm \AA} line is used in HE~1241$-$0337, 
and Mn I lines at 4451.586 and 4470.140 {\rm \AA}
were used in the case of HE~1327$-$2116. We have used the Co I 4118.770, 4121.320, 5342.695, 5483.344 {\rm \AA}
lines to estimate the cobalt abundance. We could not estimate the cobalt abundance in HE~1241$-$0337 as no good lines were 
found in the spectrum. The cobalt abundances, [Co/Fe], in the program stars 
are in the range $-$0.34 - +0.57. Copper abundance is derived using the line Cu I 5105.537 {\rm \AA}.

\subsection{Heavy elements}
\subsubsection{\textbf{The light s-process elements: Rb, Sr, Y, Zr}}
The rubidium abundance is derived from the spectral synthesis
calculation of Rb I resonance line at 7800.259 {\rm \AA}.  
The HFS components of this line are taken from \citep{Lambert_1976}. 
The other Rb I line at 7947.597 {\rm \AA} was not usable for abundance determination 
in any of the program stars. We could not derive the Rb abundance in HE~1241$-$0337 as 
this region is absent in the spectrum,  and in HE~1327$-$2116 
as we could not detect any good lines due to Rb.  
While Rb is found to be enhanced in HE~0457$-$1805 with [Rb/Fe]$\sim$1.41, 
it shows near-solar value in HE~0920$-$0506. 

\par We have estimated the Sr abundance using the spectral synthesis 
calculation of Sr I 4607.327 {\rm \AA} line. 
This line is known to be affected by NLTE effect, and the appropriate 
NLTE corrections are adopted from \cite{Bergemann_2012}. 
The Sr abundance could be determined in all the program stars except HE~1327$-$2116,  
and it is found to be enhanced with [Sr/Fe]$>$1.10. 
The NLTE corrections to the abundances of Sr for HE~0457$-$1805 and HE~1241$-$0337 are +0.47, 
and for HE~0920$-$0506 is +0.17.

\par Yttrium abundance is derived from the spectral synthesis 
calculation of Y I 6435.004 {\rm \AA} line and from the 
measured equivalent widths of several Y II lines listed in Table \ref{linelist2}. 
While we could estimate Y II abundance in all the program stars, 
Y I abundance could not be determined in HE~1241$-$0337 and HE~1327$-$2116 as we could not 
detect any good Y I lines. The Y II abundance, [Y II/Fe], ranges from 0.69 to 1.95, 
whereas Y I abundances, [Y I/Fe], are 2.78 and 0.86 respectively in HE~0457$-$1805 and 
HE~0920$-$0506. 

\par We have estimated the Zr abundance from the spectral synthesis 
calculation of  Zr I 6134.585 {\rm \AA} line in all the stars. The Zr II abundance
is derived from the measured equivalent widths of several Zr II lines in HE~1241$-$0337 and 
spectral synthesis of Zr II 5112.297 {\rm \AA} line in other three program stars. 
While Zr I abundances, [Zr I/Fe], ranges from 1.22 to 2.43, 
Zr II abundances, [Zr II/Fe], ranges from 0.55 to 2.03.

\subsubsection{\textbf{The heavy s-process elements: Ba, La, Ce, Pr, Nd}}
\begin{figure}
\centering
\includegraphics[width=\columnwidth]{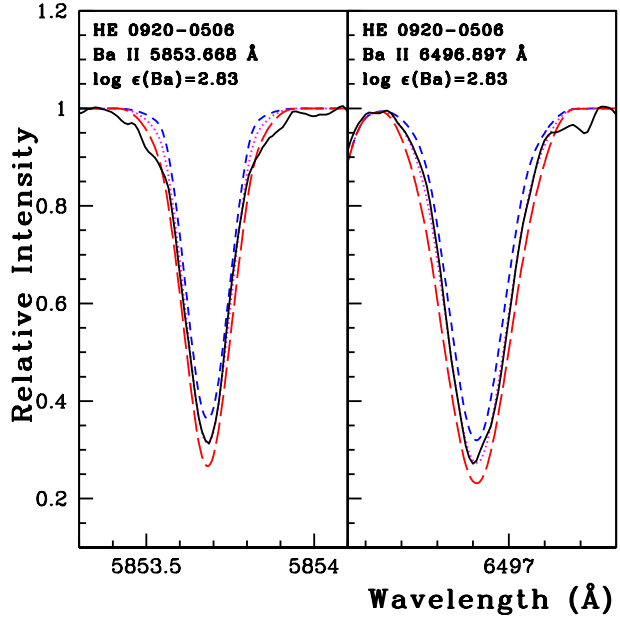}
\caption{Spectral synthesis fits of Ba II 5853.668 {\rm \AA} and 6496.897 {\rm \AA} 
lines in the star HE~0920$-$0506. Synthesized and observed spectra respectively are 
shown by dotted and solid lines. Short-dashed and long-dashed lines are the synthetic spectra 
for $\Delta$ [Ba/Fe] = $-$0.3 and +0.3 respectively.} \label{Ba_fits}
\end{figure}

In all program stars, abundance of Ba is derived from the spectral
synthesis calculation of Ba II 5853.668 {\rm \AA} line  by incorporating the hyperfine components.
The NLTE corrections to the abundances derived from this line are adopted from 
\cite{Andrievsky_2009}. The NLTE correction for HE~1241$-$0337 is +0.05, HE~1327$-$2116 
is +0.28, and 0.00 for other two stars. We could also use Ba II 6496.897 {\rm \AA} in the object 
HE~0920$-$0506. In other stars, this line is not usable. 
The Ba II 4934.076, 6141.713 {\rm \AA} lines were either strong (equivalent width$>$240 m{\rm \AA}) 
or not good to be used for abundance determination in all the objects. 
Ba is found to be enhanced in all the program stars with [Ba/Fe]$>$1.0. 
The spectral synthesis fits for Ba II 5853.668 and 6496.897 {\rm \AA} lines 
for the star HE~0920$-$0506 are shown in Figure \ref{Ba_fits}.

The La abundance is derived from the spectral synthesis calculation
of La II lines at 4748.726,  4921.776, 5259.379, and 5303.528 {\rm \AA}, 
whenever possible, by incorporating the HFS components.  
The HFS components of La II 4748.726, 5259.379 {\rm \AA} lines were not
available. All the program stars show enhancement of La with [La/Fe]$>$1, 
HE~0457$-$1805 being the most enriched in La with a value [La/Fe]$\sim$2.27. 

The abundances of Ce, Pr, and Nd are derived using the equivalent widths 
measured from several lines of the singly ionized species of these 
elements. The atomic lines used for the abundance determination of these
elements are given in Table \ref{linelist2}. All the program stars are 
found to show enhancement of [X/Fe]$>$1 for all these elements. 

We have also calculated the [ls/Fe], [hs/Fe], [hs/ls] ratios of the 
program stars. Here, ls stands for light s-process elements: Sr, Y and Zr, 
and hs stands for heavy s-process elements: Ba, La, Ce and Nd. 
We have also calculated the [s/Fe] ratio, a measure of total s-process content 
of a star, for all the program stars. Here s refers to the s-process elements 
Sr, Y, Zr, Ba, La, Ce, and Nd. The [hs/Fe] is calculated as 
([Ba/Fe] + [La/Fe] + [Ce/Fe] + [Nd/Fe])/4,  [ls/Fe] as ([Sr/Fe] + [Y/Fe] + [Zr/Fe])/3
and [s/Fe] as ([Sr/Fe] + [Y/Fe] + [Zr/Fe] + [Ba/Fe] + [La/Fe] + [Ce/Fe] + [Nd/Fe])/7.  
Values of these ratios are given in Table \ref{hs_ls_CEMP_stars}. 
As seen from the table, all our stars have high s-process content with
[s/Fe]$>1$. We could estimate the [Rb/Zr] ratio, an indicator of
neutron density at s-process site and mass of companion AGB star, for two objects. Estimated
values of this ratio  are also provided in Table \ref{hs_ls_CEMP_stars}. 
These ratios will be discussed in detail later in this paper.

{\footnotesize
\centering
\begin{table*}
\caption{Estimates of  [ls/Fe], [hs/Fe], [s/Fe], [hs/ls], [Rb/Zr]} \label{hs_ls_CEMP_stars}
\begin{tabular}{lcccccc}
\hline                       
Star name          & [Fe/H]   & [ls/Fe] & [hs/Fe]  & [s/Fe] & [hs/ls]    & [Rb/Zr]    \\ 
\hline
HE~0457$-$1805     & $-$1.98  & 1.88    & 2.38     & 2.16   & 0.50       & $-$1.02   \\
HE~0920$-$0506     & $-$0.75  & 1.29    & 1.22     & 1.15   & $-$0.07    & $-$1.29  \\
HE~1241$-$0337     & $-$2.47  & 1.20    & 1.06     & 1.31   & $-$0.14    & --      \\
HE~1327$-$2116     & $-$2.84  & 0.92    & 1.75     & 1.48   & 0.83       & --     \\
\hline
\end{tabular}
\end{table*}
}

\subsubsection{\textbf{The r-process elements: Sm, Eu}}
We have used the measured equivalent widths of several Sm II 
lines listed in Table \ref{linelist2} to derive the abundance of Sm. 
All the stars show enhanced abundance of Sm with [Sm/Fe]$\geq$0.99.
The abundance of Eu is derived from the spectral synthesis calculation 
of Eu II 6645.064 {\rm \AA} line in all the four program stars. 
We could also use Eu II line at 4129.725 {\rm \AA} in HE~0920$-$0506, 
and Eu II 6437.640 {\rm \AA} line in HE~1241$-$0337. 
The Eu II 4129.725 {\rm \AA} line is affected by NLTE, and the correction to the 
abundance derived from this line ($\sim$+0.16) is adopted from \cite{Mashonkina_2008}. 
While Eu is mildly under abundant in HE~0920$-$0506 with [Eu/Fe]$\sim$$-$0.20, 
it is moderately enhanced in HE~1241$-$0337 with [Eu/Fe]$\sim$0.52. 
The other two stars HE~0457$-$1805 and HE~1327$-$2116 are enhanced in Eu with 
[Eu/Fe] value around 1.20.

\section{ABUNDANCE UNCERTAINTIES} \label{section_uncertainty}
The uncertainty in the abundances of each element is calculated 
following the procedure in \cite{Shejeelammal_2020} using;

\begin{center}
$\sigma_{log\epsilon}^{2}$ = $\sigma_{ran}^{2}$ + $(\frac{\partial log \epsilon}{\partial T})^{2}$ $\sigma_{T\rm_{eff}}^{2}$ + $(\frac{\partial log \epsilon}{\partial log g})^{2}$ $\sigma_{log g}^{2}$ + 
  $(\frac{\partial log \epsilon}{\partial \zeta})^{2}$ $\sigma_{\zeta}^{2}$ + $(\frac{\partial log \epsilon}{\partial [Fe/H]})^{2}$ $\sigma_{[Fe/H]}^{2}$ \\    
\end{center}

Here, $\sigma_{log\epsilon}$ is the total uncertainty in the absolute 
abundance of the particular element. The $\sigma_{ran}$ is the random 
error which is calculated from the standard deviation ($\sigma_{s}$) of 
the abundances derived from N lines of the particular element using 
$\sigma_{ran}$ = $\frac{\sigma_{s}}{\sqrt{N}}$. 
The $\sigma_{T\rm_{eff}}$, $\sigma_{log g}$, $\sigma_{\zeta}$ and $\sigma_{[Fe/H]}$ 
are the typical uncertainties in the stellar atmospheric parameters, which are 
$\sigma_{T\rm_{eff}}$$\sim$$\pm$100 K, $\sigma_{log g}$$\sim$$\pm$0.2 dex, 
$\sigma_{\zeta}$$\sim$$\pm$0.2 km s$^{-1}$ and $\sigma_{[Fe/H]}$$\sim$$\pm$0.1 dex. 
The abundance uncertainty in the abundance ratio of an element, X, is calculated
from; \\ 
$\sigma_{[X/Fe]}^{2}$ = $\sigma_{X}^{2}$ + $\sigma_{Fe}^{2}$. 

We made the calculation simple by assuming that the parameters are 
independent. Hence, the uncertainties calculated here are taken to be 
the upper limits. The changes in the abundances of each element 
with the variation in different atmospheric parameters are given in Table \ref{differential_abundance}.
We have evaluated the differential abundances in the specific case of the star
HE~0920$-$0506.

{\footnotesize
\begin{table*}
\caption{Change in the abundances ($\Delta$log$\epsilon$) of different 
elemental species (of the star HE~0920$-$0506) with variations
in stellar atmospheric parameters (columns 2 - 5). Total uncertainty 
in [X/Fe] of each element is given in sixth column.}  
\label{differential_abundance}
\begin{tabular}{lccccc}
\hline                       
Element & $\Delta$T$_{eff}$  & $\Delta$log g  & $\Delta$$\zeta$       & $\Delta$[Fe/H] &  $\sigma_{[X/Fe]}$  \\
        & ($\pm$100 K)       & ($\pm$0.2 dex) & ($\pm$0.2 km s$^{-1}$) & ($\pm$0.1 dex) &                \\
\hline
C	        & $\pm$0.07	       & $\pm$0.10	    & $\mp$0.10	            & $\pm$0.10	    	  & 0.23   \\
N	        & $\pm$0.13	       & $\pm$0.03	    & 0.00	                & $\pm$0.02	     	  & 0.19   \\
O	        & 0.00	           & $\pm$0.04	    & $\pm$0.02	            & 0.00	         	  & 0.14   \\
Na I	    & $\pm$0.04	       & $\mp$0.01	    & $\mp$0.02	            & $\mp$0.01	     	  & 0.15   \\
Mg I	    & $\pm$0.07	       & $\mp$0.06	    & $\mp$0.03	            & $\pm$0.01	     	  & 0.16   \\
Si I	    & $\pm$0.04	       & 0.00	        & $\mp$0.02	            & 0.00	     	      & 0.18  \\
Ca I	    & $\pm$0.08	       & $\mp$0.04	    & $\mp$0.07	            & 0.00  	     	  & 0.17   \\
Sc II	    & $\mp$0.01	       & $\pm$0.08	    & $\mp$0.04	            & $\pm$0.02	     	  & 0.19   \\
Ti I	    & $\pm$0.11	       & $\mp$0.01	    & $\mp$0.07	            & $\mp$0.01	     	  & 0.19   \\
Ti II	    & $\mp$0.01	       & $\pm$0.06	    & $\mp$0.10	            & $\pm$0.02	     	  & 0.18    \\
V I	        & $\pm$0.13	       & $\mp$0.04      & $\mp$0.13	            & 0.00  	     	  & 0.23   \\
Cr I	    & $\pm$0.10	       & $\mp$0.03      & $\mp$0.09	            & $\mp$0.01	     	  & 0.19   \\
Cr II	    & $\mp$0.04	       & $\pm$0.08	    & $\mp$0.06	            & $\pm$0.02	    	  & 0.20   \\
Mn I	    & $\pm$0.08	       & $\mp$0.01	    & $\mp$0.08	            & $\mp$0.01	     	  & 0.17   \\
Fe I	    & $\pm$0.09	       & 0.00	        & $\mp$0.09	            & $\mp$0.01  	      & -- \\
Fe II	    & $\mp$0.05	       & $\pm$0.08	    & $\mp$0.13	            & $\pm$0.03 	      & -- \\
Co I	    & $\pm$0.11	       & 0.00	        & $\mp$0.03	            & $\mp$0.01	          & 0.17   \\
Ni I	    & $\pm$0.07	       & $\mp$0.02	    & $\mp$0.07	            & $\mp$0.01	          & 0.17   \\
Cu I	    & $\pm$0.11	       & $\mp$0.02	    & $\mp$0.16	            & $\mp$0.01	     	  & 0.23   \\
Zn I	    & $\mp$0.01	       & $\pm$0.02	    & $\mp$0.09	            & 0.00  	     	  & 0.18   \\
Rb I	    & $\pm$0.10	       & 0.00	        & $\mp$0.03	            & 0.00	       	      & 0.17   \\
Sr I	    & $\pm$0.11	       & $\mp$0.04	    & $\mp$0.18	            & 0.00  	     	  & 0.25   \\
Y I	        & $\pm$0.16	       & 0.00	        & $\mp$0.01	            & $\mp$0.01	   	      & 0.21   \\
Y II	    & $\pm$0.01	       & $\pm$0.05	    & $\mp$0.14	            & $\pm$0.02	     	  & 0.23   \\
Zr I	    & $\pm$0.12	       & 0.00  	        & $\mp$0.05	            & 0.02   	     	  & 0.18   \\
Zr II	    & $\mp$0.01	       & $\pm$0.06	    & $\mp$0.17	            & $\pm$0.01	     	  & 0.24   \\
Ba II	    & $\pm$0.05	       & $\pm$0.02	    & $\mp$0.10	            & $\pm$0.04	    	  & 0.20   \\
La II       & $\pm$0.04        & $\pm$0.03      & $\mp$0.20             & $\pm$0.02           & 0.26    \\
Ce II	    & $\pm$0.02	       & $\pm$0.08	    & $\mp$0.14	            & $\pm$0.02	     	  & 0.24   \\
Pr II	    & $\pm$0.03	       & $\pm$0.08	    & $\mp$0.03	            & $\pm$0.02	     	  & 0.19   \\
Nd II	    & $\pm$0.03	       & $\pm$0.07	    & $\mp$0.13	            & $\pm$0.02	     	  & 0.23   \\
Sm II	    & $\pm$0.03	       & $\pm$0.07	    & $\mp$0.11	            & $\pm$0.02	     	  & 0.23   \\
Eu II	    & $\mp$0.04	       & $\pm$0.07	    & $\mp$0.26	            & $\pm$0.03	      	  & 0.27   \\
\hline
\end{tabular}

\end{table*}
}

\section{Classification of the program stars} \label{section_classification}
\begin{figure}
\centering
\includegraphics[width=\columnwidth]{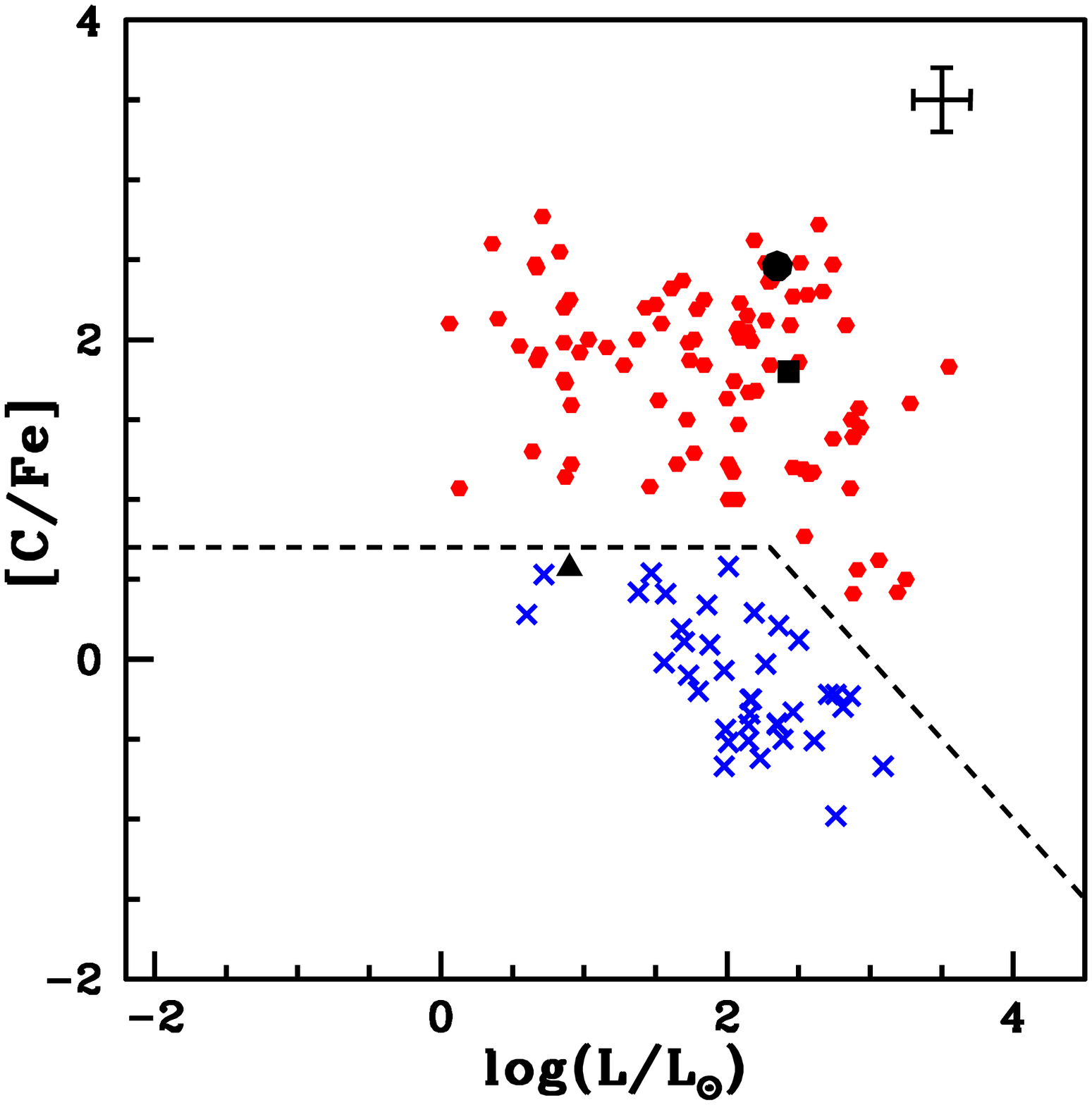}
\caption{Observed [C/Fe] ratios as a function of stellar luminosity, log(L/L$_{\odot})$.   
Red filled hexagons are CEMP stars from literature 
(\citealt{Aoki_2007} and references therein, \citealt{Goswami_2016, Purandardas_2019, Shejeelammal_2021a, Shejeelammal_2021b,  Karinkuzhi_2021, Goswami_2021, Purandardas_2021b}).
Blue crosses represent carbon-normal metal-poor stars from literature 
\citep{Aoki_2005, Aoki_2007, Cayrel_2004, Honda_2004}. 
HE~0457$-$1805 (filled square), HE~0920$-$0506 (filled triangle), 
HE~1327$-$2116 (filled circle). The dashed line separates the CEMP and non-CEMP stars. 
A representative error bar is shown at the top right corner.} \label{CEMP_Aoki}
\end{figure}

Among our stellar sample, HE~0457$-$1805, HE~1241$-$0337, and HE~1327$-$2116 are found to  
be metal-poor objects with [Fe/H]$<$$-$1, whereas HE~0920$-$0506 is moderately
metal-poor with [Fe/H]$\sim$$-$0.75.
CEMP stars are traditionally being classified as the metal-poor stars with 
[C/Fe]$>$1 \citep{Beers_2005}. This classification criteria is being 
refined and many authors use different threshold values for [C/Fe] to define CEMP stars. 
However, the carbon abundance is related to the evolutionary stages of the stars.
In the case of evolved metal-poor giant stars, the surface carbon abundance 
could be altered through CNO cycle as a result of internal mixing. 
This internal mixing processes are First Dredge-Up (FDU) on the giant branch 
and some extra mixing at the tip of RGB \citep{Charbonnel_1995, Charbonnel_1998, Gratton_2000, Shetrone_2003, Jacobson_2005, Spite_2005, Spite_2006, Aoki_2007, Placco_2014}. 
Hence, using a fixed [C/Fe] value to define CEMP stars without considering the 
evolutionary effects would be incomplete.  Here, we have adopted the 
empirical definition of \cite{Aoki_2007} to distinguish the CEMP stars 
of our sample. This definition is shown schematically in Figure \ref{CEMP_Aoki}.  
As seen from this figure, while HE~0457$-$1805 and HE~1327$-$2116 are CEMP stars, 
HE~0920$-$0506 is a non-CEMP star. From the C/O value ($>$1) of HE~0920$-$0506, we 
found that it is a CH star. We could not locate the star HE~1241$-$0337 in this diagram 
as we could not find its luminosity due to the unavailability of parallax. However, the 
higher [C/Fe] value ($\sim$2.57) of this star place it in the CEMP category. 

We have adopted the classification scheme given by \cite{Goswami_2021} to
sub-classify our CEMP stars. This classification scheme considers  
[Ba/Eu] and [La/Eu] ratios of the stars, which is given as; \\ 

\noindent-CEMP-s; [Ba/Fe]$\geq$1  \\
\indent (i) [Eu/Fe]$<$1, [Ba/Eu]$>$0 and/or [La/Eu]$>$0.5 \\
\indent (ii) [Eu/Fe]$\geq$1, [Ba/Eu]$>$1 and/or [La/Eu]$>$0.7 \\

\noindent-CEMP-r/s; [Ba/Fe]$\geq$1, [Eu/Fe]$\geq$1 \\
\indent (iii) 0$\leq$[Ba/Eu]$\leq$1 and/or 0$\leq$[La/Eu]$\leq$0.7 \\

\begin{figure}
\centering
\includegraphics[width=\columnwidth]{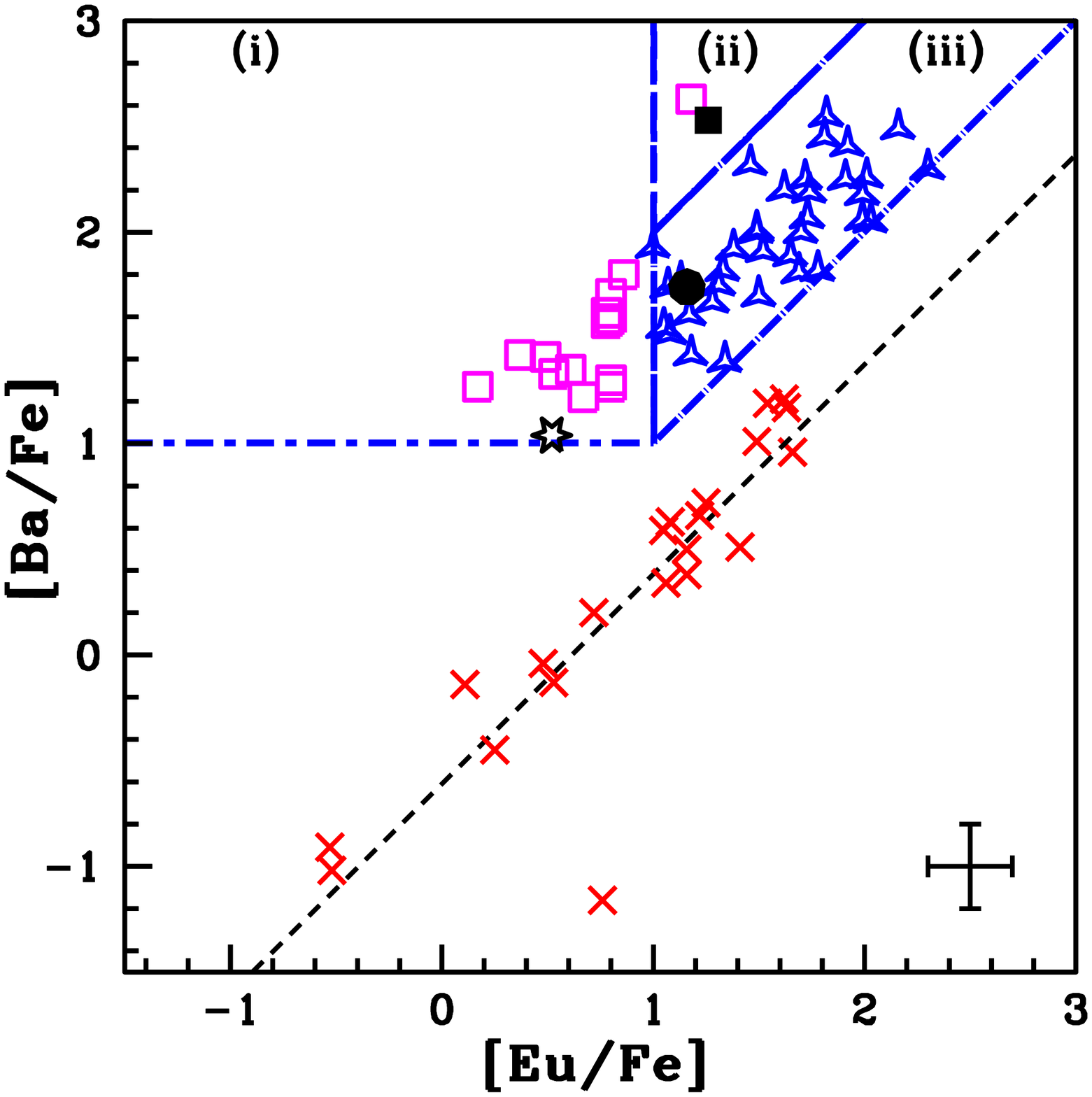}
\includegraphics[width=\columnwidth]{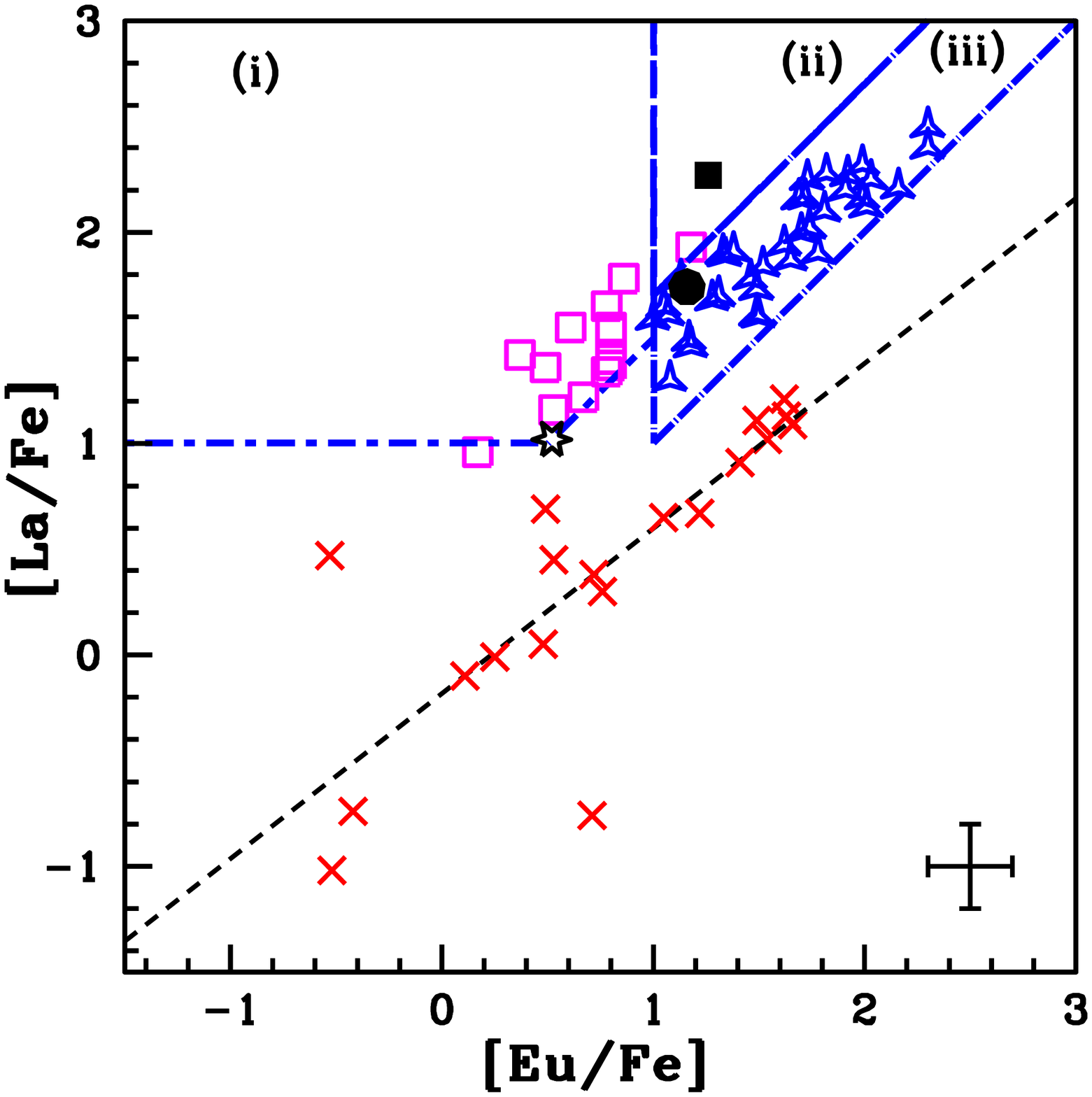}
\caption{Observed [Ba/Fe] (upper panel) and [La/Fe] (lower panel)
as functions of observed [Eu/Fe] for CEMP and r stars.   
Magenta squares are CEMP-s stars, blue starred triangles are CEMP-r/s stars, 
and red crosses are r stars (including both CEMP-r and rI/rII stars) 
from literature \citep{Masseron_2010, Shejeelammal_2021a, Shejeelammal_2021b, Karinkuzhi_2021, Goswami_2021}. 
HE~0457$-$1805 (filled square), HE~1241$-$0337 (six-sided star), 
HE~1327$-$2116 (filled circle). Region (i) and (ii) are the CEMP-s and (iii) is CEMP-r/s
star region \citep{Goswami_2021}. The least square fit to 
the abundances observed in r-stars is shown by the dashed line. 
A representative error bar is shown at the bottom right corner of each panel.
} \label{Ba_La_Eu}
\end{figure}

\noindent Figure \ref{Ba_La_Eu} schematically represents this classification. The 
values of [C/Fe], [Ba/Eu], [La/Eu] ratios observed in the program stars are given 
in Table \ref{Ba_La_Eu_ratio_CEMP_stars}. In Figure \ref{Ba_La_Eu}, the above conditions 
(i) and (ii) of CEMP-s stars are marked as region (i) and (ii) respectively, and 
CEMP-r/s condition is marked as region (iii). From this figure and Table \ref{Ba_La_Eu_ratio_CEMP_stars}, 
the object HE~0457$-$1805 and HE~1241$-$0337 are found to be CEMP-s stars (region (ii) and (i) respectively) and the 
object HE~1327$-$2116 is found to be a CEMP-r/s star (region (iii)).

{\footnotesize
\begin{table*}
\caption{The [C/Fe], [Ba/Eu] and [La/Eu] ratios in the program stars} \label{Ba_La_Eu_ratio_CEMP_stars}
\begin{tabular}{lccccccc}
\hline                       
Star name          & [Fe/H]   & [C/Fe]  & [Ba/Fe]  & [La/Fe]    & [Eu/Fe] & [Ba/Eu]    & [La/Eu]    \\ 
\hline
HE~0457$-$1805     & $-$1.98  & 1.80    & 2.53     & 2.27       & 1.26    & 1.43       & 1.01  \\
HE~0920$-$0506     & $-$0.75  & 0.57    & 1.40     & 1.25       & $-$0.20 & 1.60       & 1.45    \\
HE~1241$-$0337     & $-$2.47  & 2.57    & 1.04     & 1.02       & 0.52    & 0.52       & 0.50   \\
HE~1327$-$2116     & $-$2.84  & 2.46    & 1.74     & 1.74       & 1.16    & 0.58       & 0.58     \\
\hline
\end{tabular}
\end{table*}
}

\section{Discussion} \label{section_discussion} 
\subsection{Comparison of the observed abundance}
\par
\begin{figure}
\centering
\includegraphics[width=\columnwidth]{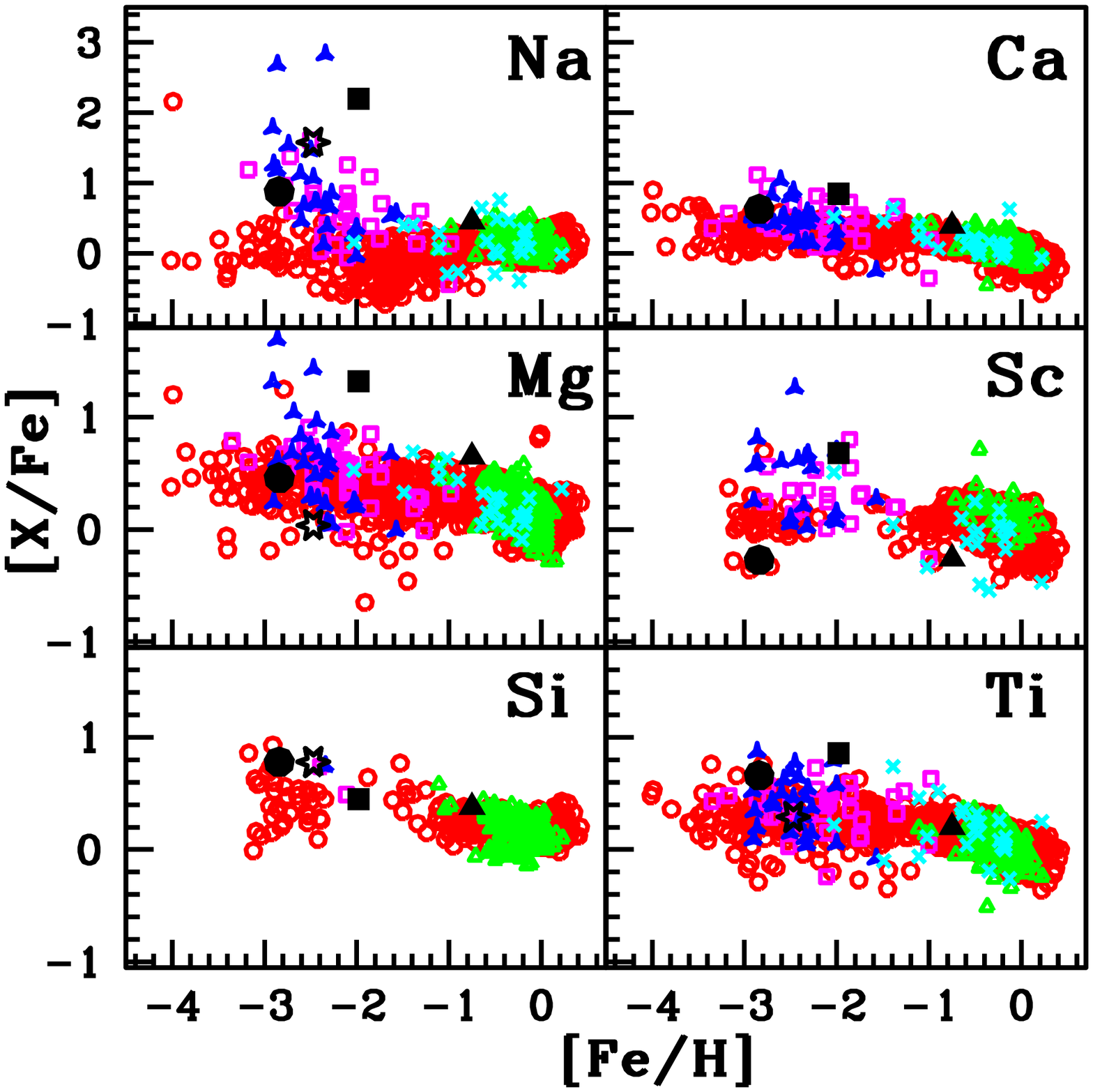}
\includegraphics[width=\columnwidth]{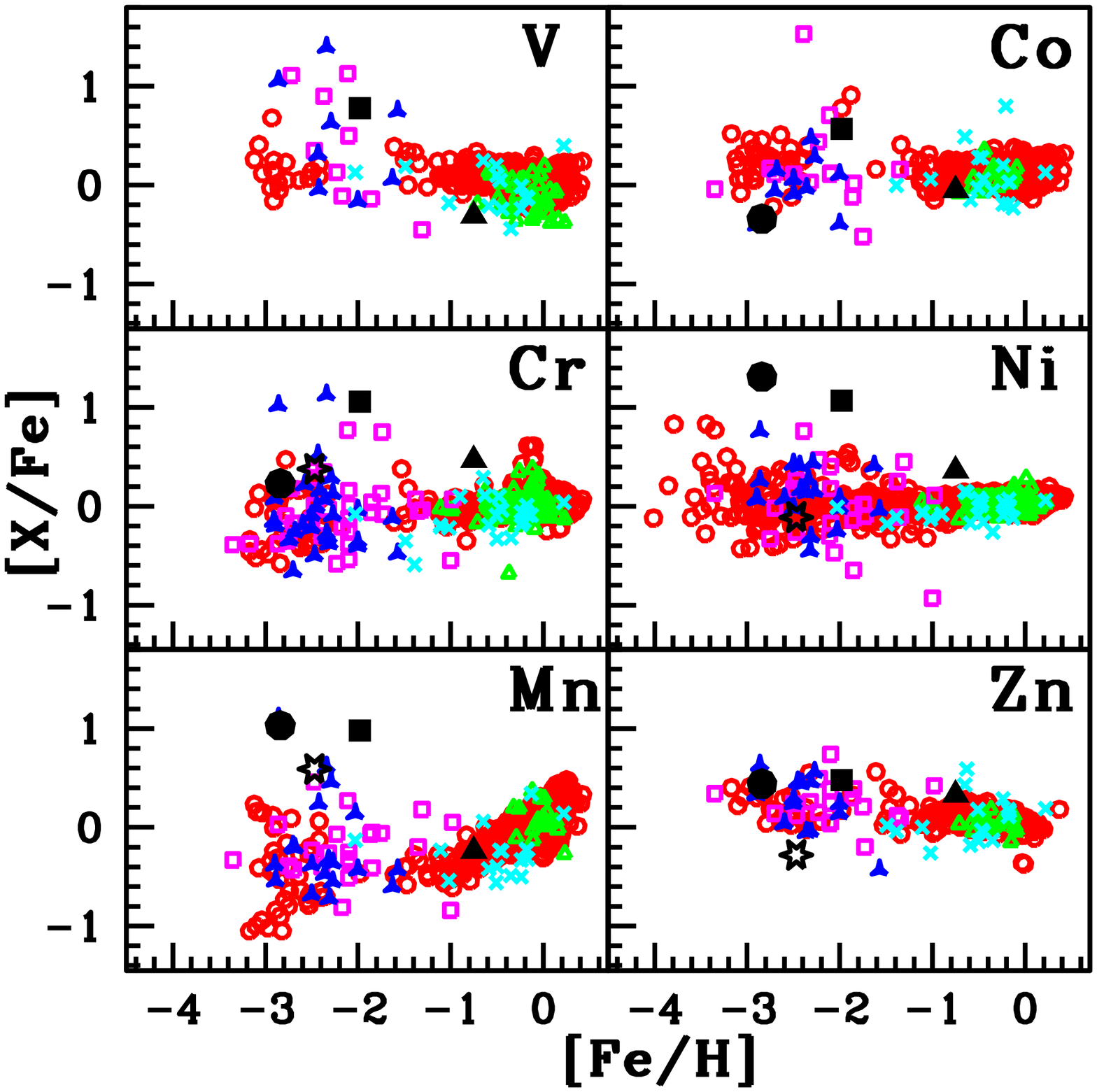}
\caption{Observed [X/Fe] ratios of the light elements in the  
program stars with respect to metallicity [Fe/H].   
Red open circles correspond to normal giants from literature 
\citep{Honda_2004, Venn_2004, Aoki_2005, Aoki_2007, Reddy_2006, Luck_2007, Hansen_2016c, Yoon_2016}.
Magenta open squares and blue starred triangles represent CEMP-s and
CEMP-r/s stars respectively from literature \citep{Masseron_2010, Purandardas_2019, Karinkuzhi_2021, Shejeelammal_2021a, Shejeelammal_2021b, Goswami_2021, Purandardas_2021b}. Cyan crosses represent CH stars from literature 
\citep{Vanture_1992, Karinkuzhi_2014, Karinkuzhi_2015, Goswami_2016, Shejeelammal_2021a, Shejeelammal_2021b}.
Green open triangles are Ba stars from literature \citep{Allen_2006a, deCastro_2016,
Yang_2016, Karinkuzhi_2018, Shejeelammal_2020}.   
HE~0457$-$1805 (filled square), HE~0920$-$0506 (filled triangle), HE~1241$-$0337 (six-sided star),
HE~1327$-$2116 (filled circle).} \label{light_elements}
\end{figure}

\begin{figure}
\centering
\includegraphics[width=\columnwidth]{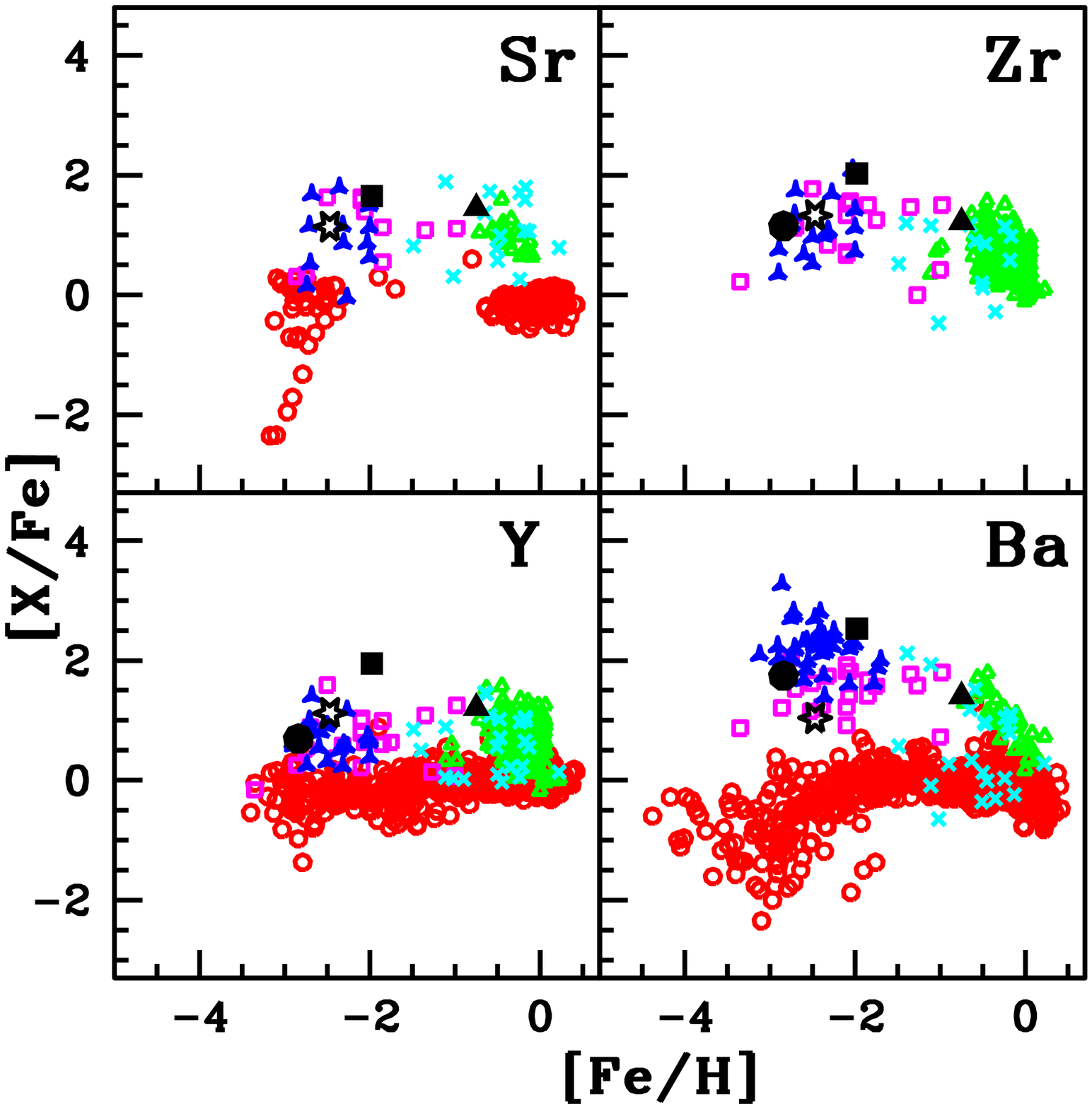}
\includegraphics[width=\columnwidth]{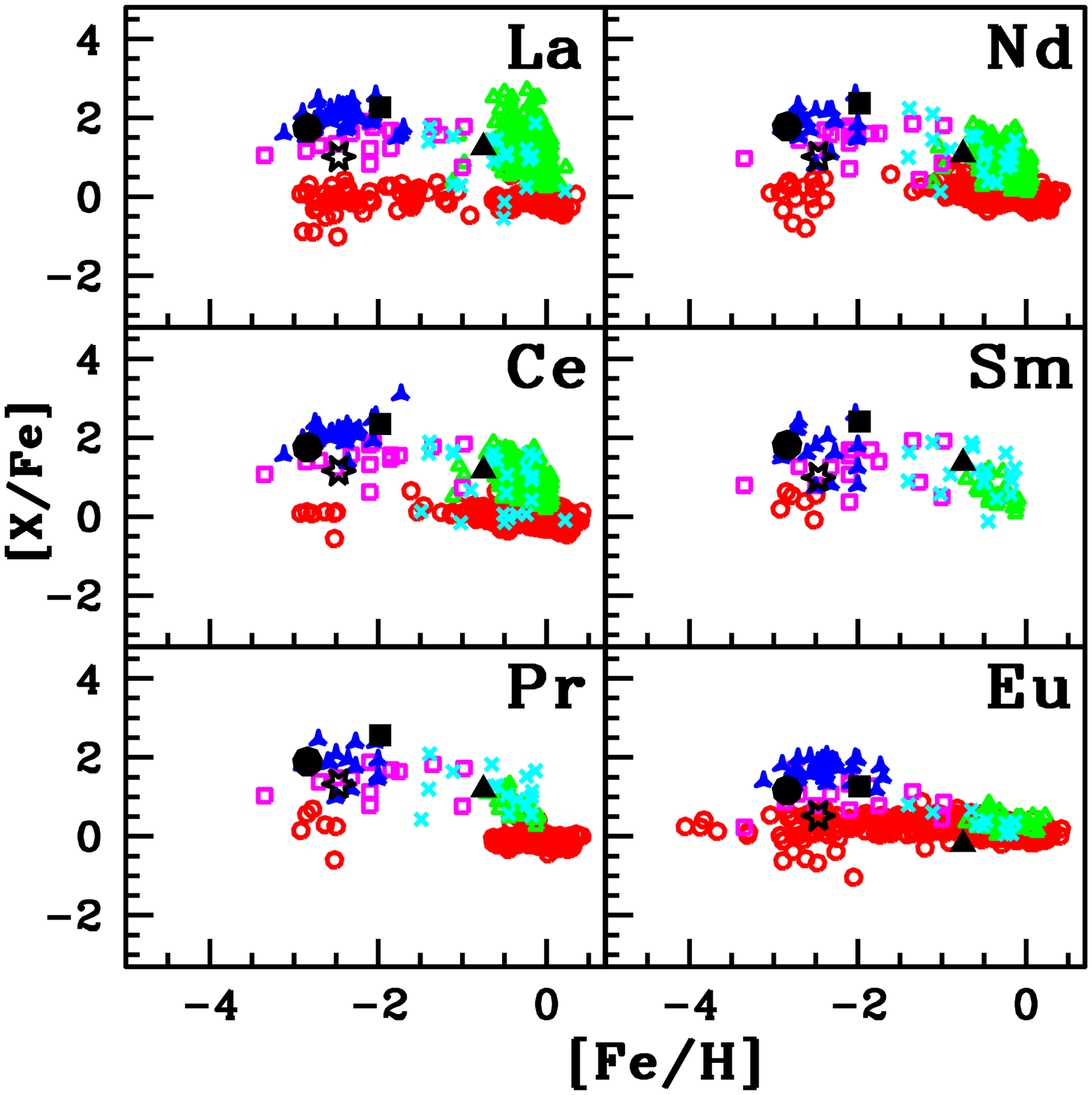}
\caption{Observed [X/Fe] ratios of the heavy elements in the  
program stars with respect to metallicity [Fe/H].   
Symbols have same meaning as in Figure \ref{light_elements}} \label{heavy_elements}
\end{figure}

 We have compared the observed abundances of light as well as heavy elements in our program stars 
with those in CH, Ba, CEMP, and normal stars from literature, and are shown in Figures \ref{light_elements} and \ref{heavy_elements}. All the s-process elements in the program stars show enhanced abundances compared to 
the normal stars. 
All the light elements except Na, Mg, V, Cr, Mn, and Ni show abundances similar to that seen in 
normal stars of the Galaxy. In HE~0457$-$1805 the elements Na, Mg, V, Cr, Mn and Ni are enhanced
when compared to normal giants. Similarly, Mn and Ni are enhanced in HE~1327$-$2116. 
Similar enhancements of Na, Mg and/or Fe-peak elements in CEMP stars were 
reported in literature \citep{Aoki_2006b, Frebel_2008, Bessel_2015, Shejeelammal_2021b, Goswami_2021, Purandardas_2021b}. 
Several sources such as faint SNe, spinstars, AGB mass transfer combined with 
metal accretion from the ISM have used to explain their abundance patterns \citep{Aoki_2006b, Frebel_2008, Bessel_2015}. 
According to \cite{Choplin_2017}, multiple sources may contribute to the formation of 
CEMP stars. This may be a possible reason for the observed enhancement of these elements. 
It is possible that the natal clouds of the CEMP stars might get polluted from multiple 
faint SNe events in the past. A study by \cite{Hartwig_2018} has presented a number of 
reliable tracers to identify whether the natal clouds of CEMP stars are mono- or 
multi-enriched. According to them, the elemental ratios [Mg/C]$<$1, [Sc/Mn]$<$0.5, 
[C/Cr]$>$0.5 or [Ca/Fe]$>$2 indicates mono-enrichment. 
We found [Ca/Fe]$<$2 in our program stars HE~0457$-$1805 and HE~1327$-$2116, which may 
indicate that these stars were formed from the gas cloud enriched by multiple SNe events.
This may also be a possible reason for the enhanced abundances of the above mentioned elements in these stars. 
Enhanced Na and/or Mg abundances are observed in a few CEMP-s and CEMP-r/s stars \citep{Bisterzo_2011, Allen_2012, Karinkuzhi_2021, Goswami_2021, Shejeelammal_2021b}. We have discussed the Na and Mg abundances 
in detail in section \ref{section_discussion_NaMg}. 

\subsection{Carbon abundance and possible origin of program stars}
Studies have shown that the absolute carbon abundances, A(C), of CEMP stars 
show bimodality, that is they tend to plateau around two distinct A(C) values 
in the A(C) - [Fe/H] diagram \citep{Spite_2013, Bonifacio_2015, Hansen_2015, Yoon_2016}.  
The CEMP-s and CEMP-r/s stars populate the high-carbon band at A(C)$\sim$7.96 
and the CEMP-no stars populate the low-carbon band at A(C)$\sim$6.28. This bimodal 
behaviour is due to the difference in the origin of carbon in the stars of these 
two bands. The carbon observed in the stars of high-carbon band (CEMP-s and CEMP-r/s) 
has an extrinsic origin: binary mass-transfer from low-mass AGB star, whereas the 
carbon in the stars of low-carbon band (CEMP-no) has an intrinsic\footnote{We used 
the term intrinsic to indicate that the observed abundance pattern of the star 
is the actual chemical imprint of the gas cloud from which the star is formed,
and not the self-enrichment of the star.} origin: pre-enrichment 
of their natal cloud by faint SNe, spinstars or metal-free massive stars 
(\citealt{Spite_2013}, \citealt{Bonifacio_2015}, \citealt{Yoon_2016} and references therein).  
This interpretation is in agreement with the results of many radial velocity monitoring 
studies of CEMP stars . Majority of the CEMP-s and CEMP-r/s stars are found 
to be binaries and a large fraction of CEMP-no stars are found to be single 
\citep{Lucatello_2005, Starkenburg_2014, Jorissen_2016a, Hansen_2016a, Hansen_2016b, Arentsen_2019b}. 
The binary fraction of CEMP-s and CEMP-r/s stars is found to be 82$\pm$10\% (18 out of 22 stars, \citealt{Hansen_2016b}) 
and of CEMP-no stars is 17$\pm$9\% (4 out of 24 stars, \citealt{Hansen_2016a}). 
 
From the compilation of a sample of CEMP stars from literature, \cite{Yoon_2016} have 
proposed that the carbon abundance A(C)$\sim$7.1 could separate binary stars from single stars in the 
A(C) - [Fe/H] diagram, though there are a few outlier stars. Majority of the binary stars
lie above this absolute carbon abundance value. A study by \cite{Arentsen_2019b} for
a sample of CEMP-no stars also confirmed the relation between high carbon abundance 
and the binarity of metal-poor stars. Hence, we have used this diagram to
get an idea about the binary status of our program stars. 

\begin{figure}
\centering
\includegraphics[width=\columnwidth]{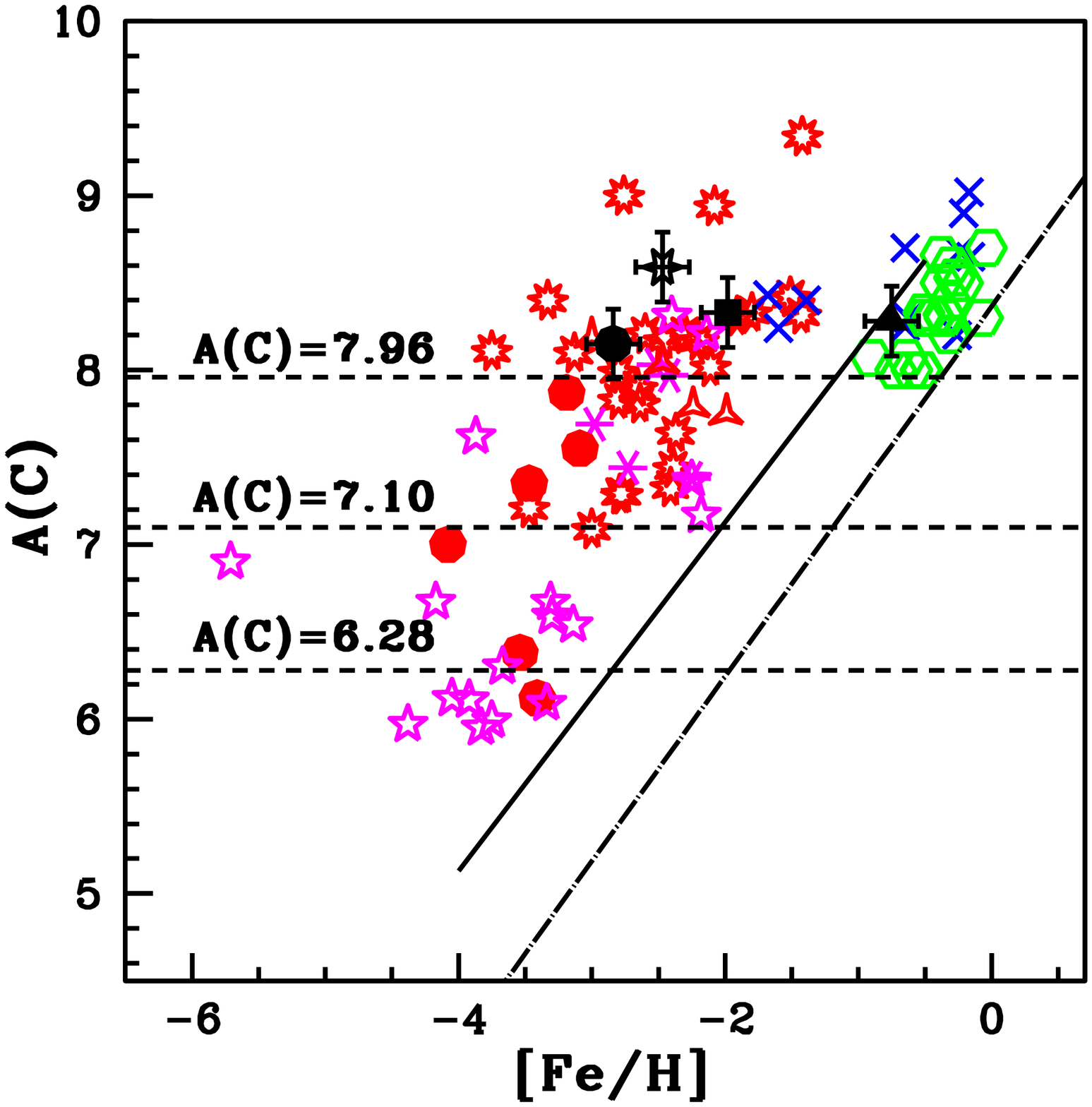}
\caption{Distribution of A(C) as a function of [Fe/H] for confirmed/likely 
binary CEMP stars from literature \citep{Yoon_2016}.     
Binary CEMP-s (red nine-sided stars), single CEMP-s (magenta six-sided crosses), 
binary CEMP-r/s (red starred triangles), binary CEMP-no (red filled circles)
and single CEMP-no (magenta five-sided stars).
All the red symbols correspond to the binary CEMP stars 
and magenta symbols to single CEMP stars.
Blue crosses represent the binary CH stars from literature 
\citep{Purandardas_2019, Karinkuzhi_2014, Karinkuzhi_2015, Luck_2017}. 
Binary Ba stars from literature \citep{Shejeelammal_2020, Karinkuzhi_2018} 
are represented by green open hexagons. 
HE~0457$-$1805 (filled square), HE~0920$-$0506 (filled triangle), 
HE~1241$-$0337 (six-sided star), 
HE~1327$-$2116 (filled circle). The black dashed line at A(C) = 7.10 separates 
binary and single stars. The high- (A(C)$\sim$7.96) and low- (A(C)$\sim$6.28) 
carbon bands of CEMP stars are also shown. The solid line is for [C/Fe] = 0.70 and 
long-dash dot line represent the solar carbon value, [C/Fe] = 0.} \label{binarity}
\end{figure}

However, before using the estimated carbon abundance for this diagram, the appropriate 
correction should be applied to it in order to account for any internal mixing. The 
internal mixing tend to alter the surface carbon abundance through CNO cycle. Hence 
the [C/N] and $^{12}$C/$^{13}$C ratio could be used as mixing diagnostics. 
Mixed stars show [C/N]$<$$-$0.6 \citep{Spite_2005} and $^{12}$C/$^{13}$C$<$10 \citep{Spite_2006}. 
However, the carbon isotopic ratio is the better indicator since the carbon and nitrogen abundances in ISM 
show larger variations \citep{Spite_2006}.
The $^{12}$C/$^{13}$C ratio is not available for HE~0920$-$0506 and HE~1241$-$0337. 
All our Program stars show [C/N]$<$$-$0.6, indicating that none of them are mixed. 
However, the object HE~1327$-$2116 shows $^{12}$C/$^{13}$C$\sim$7, which indicates 
internal mixing. The corrections to the observed absolute carbon abundances are calculated
using the public online tool developed by \cite{Placco_2014} which is available at 
\url{http://vplacco.pythonanywhere.com/}. The correction factors are 0.08, 0.03, 0.06, and 0.10
respectively for HE~0457$-$1805, HE~0920$-$0506, HE~1241$-$0337, and HE~1327$-$2116. 
The corrected A(C) values 
are used to locate the program stars in A(C) - [Fe/H] diagram, which is shown in 
Figure \ref{binarity}. We have included only those CEMP stars whose binary status is known. 
Binary CH and Ba stars from literature are also included in the figure to show their position. 

All the four program stars lie in the high-carbon band region. While HE~0920$-$0506 
lie among other CH stars, the other three stars lie among CEMP-s and CEMP-r/s stars. 
As seen from the figure, all the stars belong to the region of binary stars. 
Hence, we may expect that all the program stars are likely binaries. 
This combined with the discussion on radial velocities of the program stars in
section \ref{section_RV} may therefore point at the pollution from binary companions.

\subsection{Nature of companion AGBs of the program stars}
We have seen from the previous section that the program stars are likely binaries. 
In this section, we discuss the nature of the companion AGB stars from the analysis of 
different abundance profiles of the program stars.

\subsubsection{The [hs/ls] ratio}
The ratio of abundances of heavy s-process elements to the light s-process elements, 
the [hs/ls] ratio, is an indicator of s-process efficiency in AGB stars. 
The AGB models predict a positive value for this ratio in low-mass AGB stars ($\leq$3 M$_{\odot}$) 
and a negative value in the case of massive AGB stars ($>$3 M$_{\odot}$) 
\citep{Busso_2001, Goriely_2005, Karakas_2010, Karakas_2012, vanRaai_2012, Karakas_2014}. 
As seen from Table \ref{hs_ls_CEMP_stars}, while the stars HE~0457$-$1805 and HE~1327$-$2116 
show positive values for [hs/ls] ratio (0.50 and 0.83 respectively), HE~0920$-$0506 and HE~1241$-$0337 show 
negative values, $\sim$$-$0.07 and $-$0.14 respectively. This implies that light s-process 
elements are produces largely compared
to the heavy s-elements, characteristic of massive AGB stars. However, negative values for this 
ratio have been reported in literature for a few stars whose companions are found to be low-mass
AGB stars from the comparison of observed abundances and AGB nucleosynthesis models 
\citep{Aoki_2002, Bisterzo_2011, Hansen_2019, Shejeelammal_2021b}. 
In massive AGB stars, the nitrogen is enhanced compared to carbon as a result of 
Hot-Bottom Burning (HBB) (e.g. \citealt{McSaveney_2007, Johnson_2007, Karakas_2014}). 
However, in HE~0920$-$0506 and HE~1241$-$0337 the observed [C/N] values are $\sim$0.12 and 1.55 
respectively. 
This observed C and N abundances in HE~0920$-$0506 and HE~1241$-$0337 may therefore rule out the 
possibility of massive AGB companion.

\subsubsection{Na, Mg, and heavy elements} \label{section_discussion_NaMg}
In massive AGB stars with HBB, besides N, sodium is also expected to produce in abundance 
through Ne - Na cycle (e.g. \citealt{Mowlavi_1999b}). The $^{22}$Ne($\alpha$, n)$^{25}$Mg 
in the massive AGB stars results in the enhanced abundance of Mg too (e.g. \citealt{Karakas_2003}).
Among our program stars, three stars are enhanced in Na: HE~1327$-$2116 with [Na/Fe]$\sim$0.88, 
HE~1241$-$0337 with [Na/Fe]$\sim$1.58, and HE~0457$-$1805 with [Na/Fe]$\sim$2.20. The star HE~0457$-$1805 
is also enhanced in magnesium with [Mg/Fe]$\sim$1.32. However, the [hs/ls] ratio ($\sim$0.50) 
and [C/N] value ($\sim$0.13) of HE~0457$-$1805 and the higher [C/N] ratio of HE~1241$-$0337 ($\sim$1.55) 
discard the possibility of massive AGB companion.  
A few other studies have already reported such higher enhancement of Na and/or Mg in CEMP stars, 
for instance, CS~29528$-$028, [Fe/H]$\sim$$-$2.86, [Na/Fe]$\sim$2.68, [Mg/Fe]$\sim$1.69 \citep{Aoki_2007},  
SDSS~1707+58, [Fe/H]$\sim$$-$2.52, [Na/Fe]$\sim$2.71, [Mg/Fe]$\sim$1.13 \citep{Aoki_2008} 
and HE~1304$-$2111, [Fe/H]$\sim$$-$2.34, [Na/Fe]$\sim$2.83 \citep{Shejeelammal_2021b}. 
The star HE~1304$-$2111 is found to have a low-mass AGB companion from a detailed 
abundance analysis \citep{Shejeelammal_2021b}. 
\cite{Bisterzo_2011} have analyzed the observed abundances in a sample of 
CEMP-s stars ($\sim$100 objects) from literature using the AGB models of \cite{Bisterzo_2010}. 
These AGB models have considered $^{23}$Na and $^{24}$Mg to be primary, 
produced through $^{22}$Ne(n, $\gamma$)$^{23}$Ne($\beta^{-}\nu$)$^{23}$Na and 
$^{23}$Na(n, $\gamma$)$^{24}$Na($\beta^{-}\nu$)$^{24}$Mg. If the $^{23}$Na produced is primary,
a larger amount of it would be expected \citep{Mowlavi_1999b, Gallino_2006}. 
These models predicted high Na abundances at low-metallicities. 
The analysis of \cite{Bisterzo_2011} have shown that the higher Na abundances of 
CS~29528$-$028 and SDSS~1707+58 could be reproduced with AGB models of 
M$\rm_{AGB}^{ini}\sim$1.5 M$_{\odot}$. Their higher [ls, hs/Fe] values could be reproduced with
M$\rm_{AGB}^{ini}\sim$2.0 M$_{\odot}$ models. The entire observed abundance pattern could not be 
reproduced with the same AGB model.  

In AGB models, considering the Partial Mixing (PM) of protons, 
$^{23}$Na is produced efficiently, that is almost fifty times higher than
that produced in the H-burning shell \citep{Goriely_2000}.
In such a scenario, Na is produced in the inter-shell,
in  a region above the region of s-process, and a correlation between
[Na/Fe] and [s/Fe] is expected. This will entirely depend on the 
extent of partial mixing zone. However, these models could not explain the 
least Na-enriched CEMP-s and most Na-enriched CEMP-r/s stars (Figure 19 of \citealt{Karinkuzhi_2021}).
Better models with improved simulations are needed to explain this.

\subsubsection{The [Rb/Zr] ratio}
\begin{figure}
\centering
\includegraphics[width=\columnwidth]{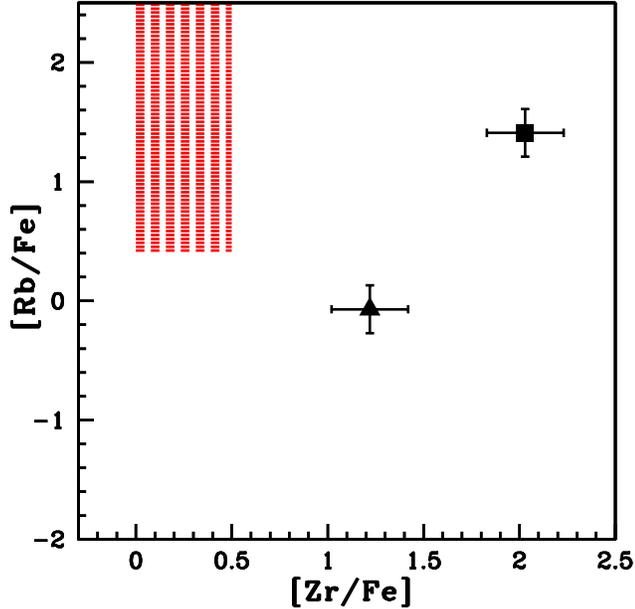}
\caption{The observed [Rb/Fe] and [Zr/Fe] in the program stars.
HE~0457$-$1805 (filled square), HE~0920$-$0506 (filled triangle). 
The shaded region is the observed ranges of [Rb/Fe] and [Zr/Fe] 
in intermediate-mass AGB stars of the Galaxy and the Magellanic 
Clouds \citep{vanRaai_2012}.} \label{Rb_Zr}
\end{figure}

Based on the C, N abundances, [hs/ls] ratio, Na and Mg abundances in the program stars,
we have ruled out the possibility of massive-AGB companions for our program stars. 
We have tried to establish an upper limit to the companion's mass from the neutron-density 
dependent [Rb/Zr] ratio. Massive AGB stars (M$\geq$ 4 M$_{\odot}$) are 
characterized by positive values of [Rb/Zr] ratio and low-mass AGB stars (M$\leq$ 3 M$_{\odot}$) 
by negative values of [Rb/Zr] ratio \citep{Abia_2001, vanRaai_2012, Karakas_2012}. 
Detailed discussion on [Rb/Zr] ratio is presented in \cite{Shejeelammal_2020}. 
We could estimate the value of this ratio in HE~0457$-$1805 and HE~0920$-$0506. 
Both the stars show negative value for this ratio ($<$$-$1, Table \ref{hs_ls_CEMP_stars}). 
A comparison of observed Rb and Zr abundances of the program stars with their counterparts
in intermediate-mass AGB stars of the Galaxy and Magellanic Clouds is shown in Figure \ref{Rb_Zr}. 
The Rb and Zr abundances of AGB stars are taken from \cite{vanRaai_2012}. 
It is clear from the figure that the [Rb/Fe] and [Zr/Fe] observed in the program stars 
do not match closely with their counterparts observed in the intermediate-mass AGB stars.
This confirms the low-mass nature of the companion AGB stars. 

\subsection{Parametric model based analysis} \label{section_parameric_model}
Our analysis based on different abundance ratios of the program stars have confirmed the 
pollution from low-mass AGB companions. To corroborate our results, we have conducted a parametric 
model based analysis for our sample. The observed abundances of neutron-capture elements in 
our program stars are compared with the predictions of stellar nucleosynthesis models appropriate
for each class of star, which we will discuss here. 

\subsubsection{CEMP-s and CH stars} 
\begin{figure}
\centering
\includegraphics[width=\columnwidth]{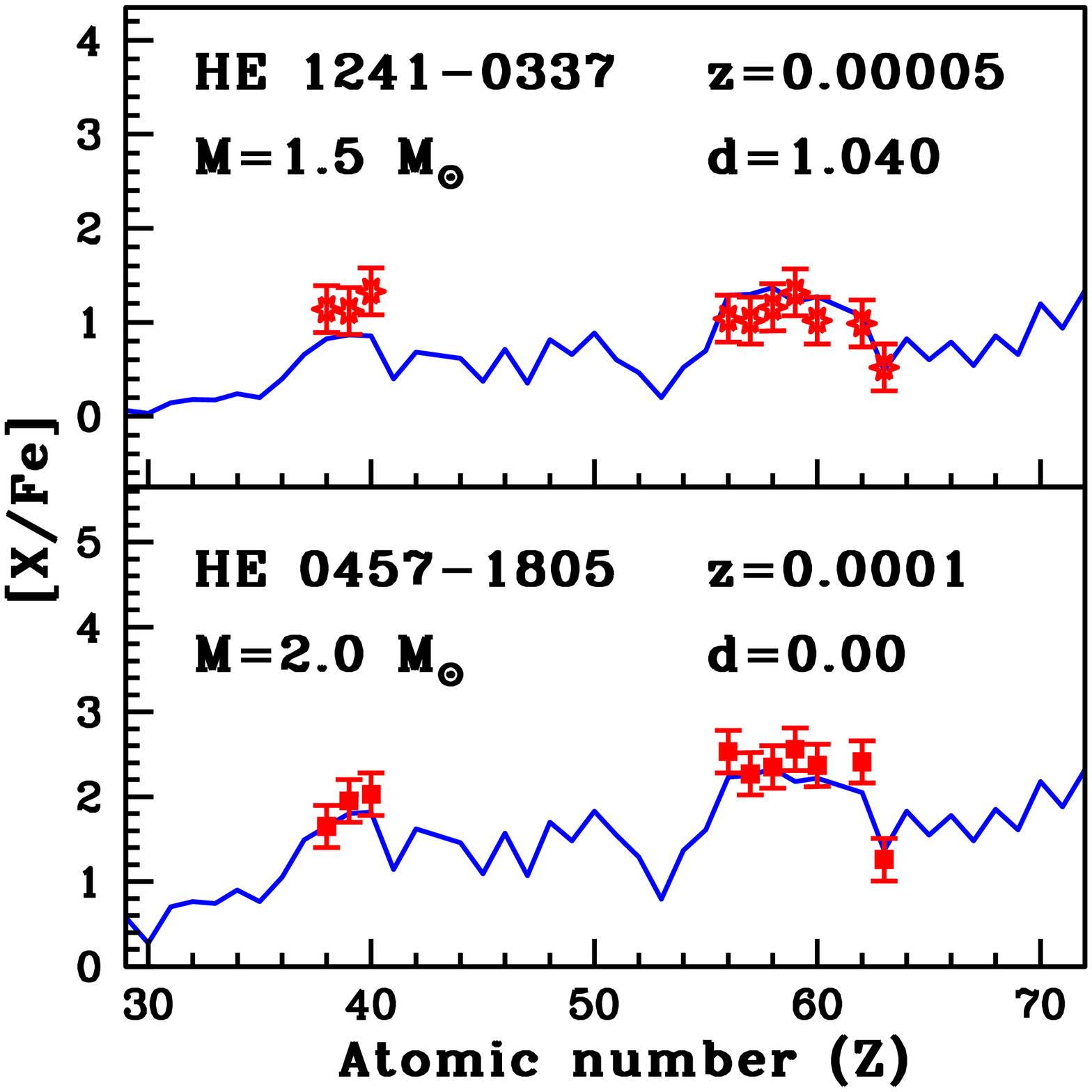}
\includegraphics[width=\columnwidth]{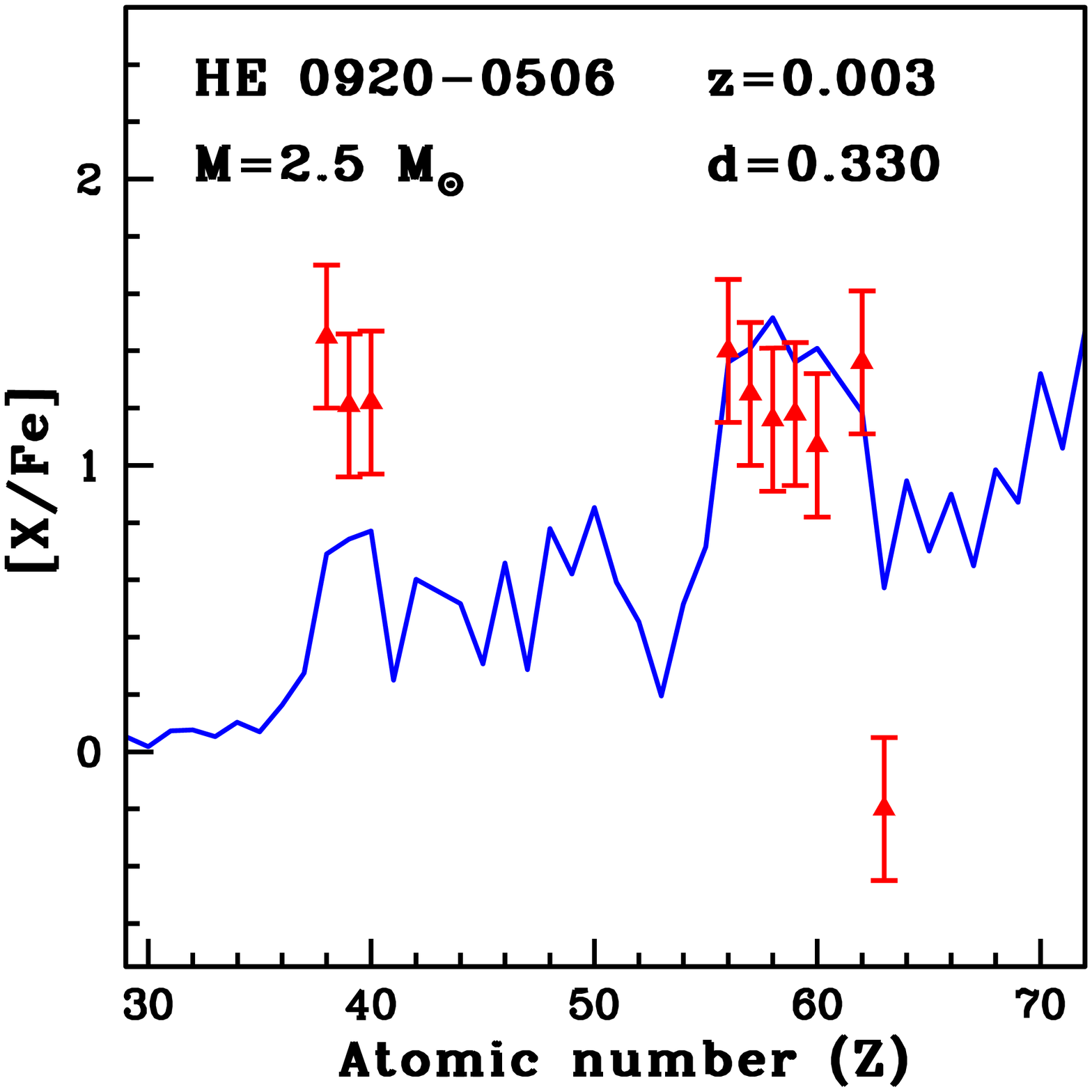}
\caption{The parametric model fits for the CEMP-s stars, HE~0457$-$1805 and 
HE~1241$-$0337 (upper panel), and CH star HE~0920$-$0506 (lower panel) using 
the FRUITY models. Solid curves represent the best fit for the parametric 
model function. The points with error bars indicate the observed abundances. 
In the figure, z is the metallicity of the FRUITY model used, and d is the 
dilution factor, a free parameter which is varied to find the best fit between 
the model and the observed abundance.} 
\label{parametric_CH_CEMP_s}
\end{figure}

For the objects HE~0457$-$1805, HE~1241$-$0337 (CEMP-s), and HE~0920$-$0506 (CH), the observed abundances are 
compared with the predicted abundances for s-process in AGB stars from 
FRUITY (FRANEC Repository of Updated Isotopic Tables \& Yields) models \citep{Cristallo_2009, Cristallo_2011, Cristallo_2015b}. 
The FRUITY models are available for the range of metallicities from z = 0.000020 to 0.020 
and for the mass range from 1.3 - 6.0 M$_{\odot}$, which is publicly accessible at \url{http://fruity.oa-teramo.inaf.it/}.  
We have derived the mass of AGB companions of these three stars by minimizing the $\chi^{2}$ 
between observed and predicted abundances, using the parametric model function of \cite{Husti_2009}. 
The dilution experienced by the material on the surface of the program stars is also
incorporated in the calculation. The detailed procedure is discussed in \cite{Shejeelammal_2020}. 
The best fits obtained for the observed abundance patterns in these stars are shown in Figure \ref{parametric_CH_CEMP_s}. 
The former AGB companions of HE~0457$-$1805, HE~0920$-$0506, and HE~1241$-$0337 are found to have 
masses 2.0, 2.5, 1.5 M$_{\odot}$ respectively.

\subsubsection{CEMP-r/s star} 
\begin{figure}
\centering
\includegraphics[width=\columnwidth]{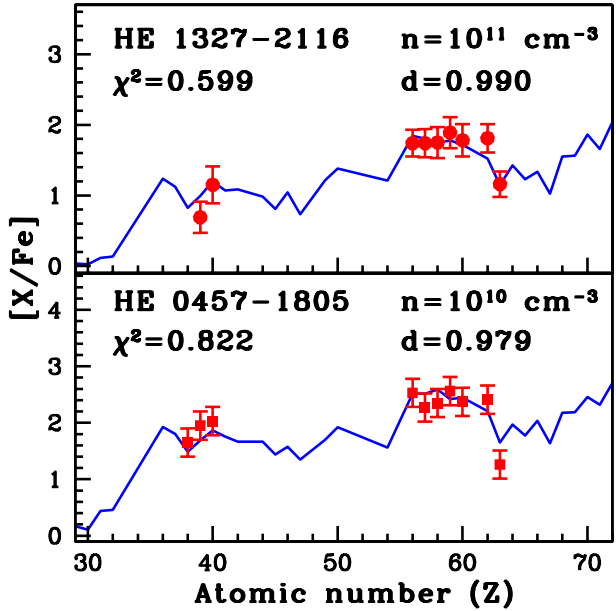}
\caption{The parametric model fits for HE~0457$-$1805 (CEMP-s) and HE~1327$-$2116 (CEMP-r/s) 
using the i-process model yields \citep{Hampel_2016}. 
Solid curves represent the best fit for the parametric model function.
The points with error bars indicate the observed abundances.} 
\label{parametric_CEMP}
\end{figure}

Among the number of scenarios proposed for the origin of CEMP-r/s stars (see for e.g. 
\citealt{Jonsell_2006} and references therein), several studies have shown that i-process 
in low-mass, low-metallicity AGB stars is a promising mechanism to explain their 
origin (e.g. \citealt{Hampel_2016, Hampel_2019, Karinkuzhi_2021, Goswami_2021, Shejeelammal_2021a, Shejeelammal_2021b, Purandardas_2021b}). Here, we have compared the observed abundances in our CEMP-r/s star 
HE~1327$-$2116 with the i-process model yields of \cite{Hampel_2016}. The analysis is performed
for a range of neutron densities from n$\sim$10$^{9}$ to 10$^{15}$ cm$^{-3}$ using the 
parametric model function given in \cite{Hampel_2016}, following the procedure discussed in 
\cite{Shejeelammal_2021b}. The best fit obtained for the observed abundance pattern in 
HE~1327$-$2116 is shown in Figure \ref{parametric_CEMP}. We found that the abundance pattern in this star
could be reproduced with the i-process models of neutron density n$\sim$10$^{11}$ cm$^{-3}$. 
We also found that the abundance pattern of the CEMP-s star HE~0457$-$1805 of our sample 
could be reproduced with a neutron density n$\sim$10$^{10}$ cm$^{-3}$, indicating the s-process. 
The parametric model fit for this star is also shown in Figure \ref{parametric_CEMP}. 

\subsection{Discussion on individual stars} \label{section_individual_stars}

\noindent\textbf{HE~0457$-$1805:} 
This object is listed among the faint high latitude carbon stars identified from 
the Hamburg/ESO Survey \citep{Christlieb_2001a}. \cite{Goswami_2005} identified HE~0457$-$1805 
as a potential CH star candidate from low-resolution spectroscopic analysis. 
The object HE~0457$-$1805 is found to be a metal-poor star with [Fe/H]$\sim-$1.98. 
Our analysis shows that HE~0457$-$1805 is a CEMP-s star. 
\cite{Kennedy_2011} have reported the stellar atmospheric parameters and C and O abundances
of this object from medium-resolution (R$\sim$3000) optical and NIR spectroscopic analysis. 
Our estimate of atmospheric parameters are in agreement with \cite{Kennedy_2011} 
except for [Fe/H] which is 0.5 dex higher than our estimate. They have reported a [C/Fe] value which is 
$\sim$0.81 dex lower than our estimate.  \\

\noindent\textbf{HE~0920$-$0506, HE~1327$-$2116:}
These objects belong to the sample of bright metal-poor candidates selected from HES 
survey \citep{Frebel_2006} through medium-resolution (R$\sim$2000) spectroscopic analysis.
\cite{Frebel_2006} have derived the metallicities of these objects from the Ca II K 3933 {\rm \AA}
line using the calibration equation of \cite{Beers_1999}. Within the error limits, these estimates
agrees with our metallicity values. A more refined analysis of \cite{Beers_2017} using the 
same medium-resolution spectra as \cite{Frebel_2006} had reported the stellar atmospheric parameters
(T$\rm_{eff}$, log g, [Fe/H])$\sim$(5291 K, 2.99, $-$1.39) and (4868 K, 0.55, $-$3.48) respectively
for HE~0920$-$0506 and HE~1327$-$2116. In the case of HE~0920$-$0506, while our estimate of 
temperature agrees with their temperature estimate, our estimates of surface gravity 
is $\sim$0.34 dex lower and metallicity is $\sim$0.64 dex higher. 
In the case of HE~1327$-$2116, the temperature estimates are in agreement, whereas 
our estimates of metallicity and surface gravity respectively are 0.64 and 0.95 dex higher than 
their estimates. These differences in the atmospheric parameters may be due to the
difference in the method adopted for their determination and also due to the different resolution of the
spectra used. While \cite{Frebel_2006} reported the carbon abundances [C/Fe]$\sim$0.09 and 1.11 
respectively for HE~0920$-$0506 and HE~1327$-$2116, \cite{Beers_2017} reported [C/Fe]$\sim$1.19 and 2.64.
Our estimates of carbon abundance for these two stars are [C/Fe]$\sim$0.57 and 2.46 respectively. 
From our analysis, we found that HE~0920$-$0506 is a CH star and HE~1327$-$2116 is a CEMP-r/s star. \\

\noindent\textbf{HE~1241$-$0337:} This object belongs to the catalog of faint high latitude 
carbon stars identified from the Hamburg/ESO Survey \citep{Christlieb_2001a}. 
We present the first-time abundance analysis for this object. 
We found from our analysis that this object is CEMP-s star with [Fe/H]$\sim$$-$2.47 
and [C/Fe]$\sim$2.57. This object is found to be a high-radial velocity object.
Parametric model based analysis shows that the former AGB companion of 
HE~1241$-$0337 is low-mass star with mass $\sim$ 1.5 M$_{\odot}$.

\section{conclusions} \label{section_conclusion}
The results of a detailed spectroscopic analysis of four 
potential CH/CEMP star candidates selected from the Hamburg/ESO Survey 
are presented. We present the first ever abundance analysis for the object HE~1241$-$0337. 
Although the abundances of Fe and C (and O for HE~0920$-$0506) 
derived from medium-resolution spectra are available in the literature, we 
present for the first time a high-resolution spectroscopic analysis for the objects 
HE~0457$-$1805, HE~0920$-$0506, and HE~1327$-$2116. 
We have estimated the stellar atmospheric parameters as well as the abundances 
of twenty-eight elements along with the carbon isotopic ratio. 

Our analysis has shown that the objects HE~0457$-$1805 and HE~1241$-$0337 are CEMP-s stars, 
HE~0920$-$0506 is a CH star, and HE~1327$-$2116 is a CEMP-r/s star. 
The radial velocity estimate shows that HE~0457$-$1805 and HE~0920$-$0506 are 
low-radial velocity objects ($<$ 100 km s$^{-1}$), whereas HE~1241$-$0337 and HE~1327$-$2116 
are high-radial velocity ($>$ 100 km s$^{-1}$) objects. 
While HE~0457$-$1805 is a confirmed binary, the difference noted in 
the radial velocity of HE~0920$-$0506 from literature value may indicate 
that it is likely a binary. Our analysis based on absolute carbon abundance 
has revealed that all these stars belong to the region of binary stars. 

The positions of program stars on the A(C) - [Fe/H] diagram suggest 
that the observed enhancement in the abundances of the neutron-capture 
elements may be due to the pollution from binary companions. 
We have investigated the nature of the companion AGB stars 
of our program stars using several diagnostics such as C, N, Na, and Mg 
abundances, [hs/ls] and [Rb/Zr] ratios. From the analysis based on these
diagnostics, we have found that none of the program stars are polluted by 
massive AGB stars. Our analysis based on different abundance ratios and abundance 
profiles confirmed the low-mass AGB companions of the program stars. 

We have carried out a parametric model based analysis for the CEMP-s (HE~0457$-$1805 and HE~1241$-$0337) and
CH (HE~0920$-$0506) stars in our sample. A comparison of the observed abundances in 
these stars with the predictions from the FRUITY models of AGB stars by means of a dilution 
factor incorporated parametric model function confirmed that the AGB stars that polluted them 
were low-mass stars with masses M $\leq$ 2.5 M$_{\odot}$. 

The observed abundance pattern in the CEMP-r/s star of our sample, HE~1327$-$2116, 
is well reproduced with the i-process parametric models in low-mass, low-metallicity 
AGB stars. The neutron density responsible for its observed abundance is found to be 
n$\sim$10$^{11}$ cm$^{-3}$. Our analysis for HE~0457$-$1805 using these same models 
has shown that the neutron density n$\sim$10$^{10}$ cm$^{-3}$ in low-mass AGB stars could reproduce 
its observed abundance. Thus, the parametric model based study for the program stars 
corroborates the results we have obtained from the abundance profile analysis.

 \section{ACKNOWLEDGMENT}
Funding from the DST SERB project No. EMR/2016/005283 is gratefully 
acknowledged. We are thankful to the referee for useful 
comments and suggestions.
We are thankful to Melanie Hampel for providing us with 
the i-process yields in the form of number fractions. We thank  Alain Jorissen for sharing the 
HERMES spectra used in this study. The HERMES spectrograph is supported by the Fund for
Scientific Research of Flanders (FWO), the Research Council of K.U.Leuven,
the Fonds National de la Recherche Scientifique (F.R.S.- FNRS), Belgium, the
Royal Observatory of Belgium, the Observatoire de Gen\'eve, Switzerland and the
Th\"uringer Landessternwarte Tautenburg, Germany.
This work made use of the SIMBAD astronomical database, operated
at CDS, Strasbourg, France, and the NASA ADS, USA.
This work has made use of data from the European Space Agency (ESA) 
mission Gaia (\url{https://www.cosmos.esa.int/gaia}), processed by the Gaia 
Data Processing and Analysis Consortium 
(DPAC, \url{https://www.cosmos.esa.int/web/gaia/dpac/consortium}). \\

\noindent
{\bf Data Availability}\\
The data underlying this article will be shared on reasonable request to the authors.

\bibliographystyle{mnras}
\bibliography{Bibliography}

\appendix
\restartappendixnumbering
\section{Line list} \label{section_linelist}
The lines used to derive the elemental abundances are listed in Tables \ref{linelist1} - \ref{linelist3}. 
{\footnotesize
\begin{table*}
\caption{Equivalent widths (in m\r{A}) of Fe lines used for deriving 
atmospheric parameters.} \label{linelist1}
\resizebox{\textwidth}{!}{\begin{tabular}{lccccccc}
\hline                       
Wavelength(\r{A}) & El & $E_{low}$(eV) & log gf & HE~0457$-$1805 & HE~0920$-$0506 & HE~1241$-$0337 & HE~1327$-$2116   \\ 
\hline 
4132.9		& Fe I  &	2.85	&	$-$1.01		&	-		    &	76.1(6.73)	& -             &	-				\\
4337.05		&		&	1.56	&	$-$1.7		&	145.6(5.42)	&	-		    & -             &	-				\\
4422.568	&		&	2.845	&	$-$1.11		&	-		    &	74.8(6.69)	& -             &	-				\\
4531.626	&		&	3.211	&	$-$2.511	&	-	    	&	-		    & -             &	83.4(4.68)			\\
4547.847	&		&	3.546	&	$-$0.780	&	-	    	&	62.6(6.65)	& -             &	-				\\
4625.045	&		&	3.241	&	$-$1.34		&	61.9(5.49)	&	58.0(6.73)	& -             &	-				\\
4630.120    &       &   2.277   &   $-$2.580    &   -           &   -           & 51.1(5.15)    &   -              \\
4733.591	&		&	1.484	&	$-$2.71		&	127.0(5.66)	&	-	    	& -			    &   -         	\\
4924.770    &       &   2.277   &   $-$2.220    &  -            &   -           & 67.3(4.95)    &   -          \\
4969.917	&		&	4.216	&	$-$0.71		&  -		    &	45.3(6.70)	&	-			&   -	\\
4985.253    &       &   3.926   &   $-$0.440    &  -            & -             & 35.4(4.90)    &   -           \\ 
5005.711	&		&	3.883	&	$-$0.18		&	82.8(5.42)	&	81.3(6.88)	&	-			&   -    	\\
5006.119	&		&	2.833	&	$-$0.61		&	143.9(5.63)	&	-	        &	-			&   -  	\\
5044.211    &       &   2.849   &   $-$2.150    &  -            & -             & 23.9(5.01)    &   -           \\ 
5049.820    &       &   2.277   &   $-$1.420    &  -            & -             & 150.7(5.12)   &   -           \\ 
5079.223	&		&	2.2	    &	$-$2.067	&	115.6(5.65)	&	72.5(6.71)	& -             &	27.3(4.72)		\\
5127.359	&		&	0.915	&	$-$3.307	&	-	    	&	80.9(6.78)	&	-			&   -	\\
5151.911    &       &   1.010   &   $-$3.320    &  -            & -             & 137.6(5.05)   &   -           \\ 
5195.468	&		&	4.22	&	$-$0.02		&	71.2(5.49)	&	74.2(6.83)	&	-			&   -   	\\
5198.711	&		&	2.222	&	$-$2.135	&	-	    	&	-		    &	-           &   20.6(4.67)			\\
5215.179	&		&	3.266	&	$-$0.933	&	-		    &	-		    &	-           &   20.7(4.69)		\\
5226.862	&		&	3.038	&	$-$0.667	&	115.7(5.34)	&	92.0(6.76)	&	-           &   45.4(4.57)		\\
5247.05		&		&	0.087	&	$-$4.946	&	-		    &	53.8(6.70)	&	-			&   -   	\\
5307.37		&		&	1.61	&	$-$2.192	&	138.7(5.32)	&	-	    	&	-			&   -   	\\
5339.93		&		&	3.27	&	$-$0.680	&	117.8(5.66)	&	-	    	&	-			&   -   	\\
5379.574	&		&	3.694	&	$-$1.480	&	-	    	&	40.8(6.78)	&	-			&   -   	\\
5393.17		&		&	3.24	&	$-$0.720	&	-		    &	-		    &	-           &   24.2(4.71)		\\
5569.62		&		&	3.42	&	$-$0.490	&	-		    &	-		    &	-           &   31.2(4.66)			\\
5576.089    &       &   3.428   &   $-$1.000    &   -           &   -           &   68.88(5.22) &   -           \\ 
5701.544    &       &   2.557   &   $-$2.220    &   -           &   -           &   49.1(4.94)  &   -           \\ 
5809.22		&		&	3.88	&	$-$1.690	&	-		    &	27.4(6.74)	&	-			&   -   	\\
5862.357	&		&	4.549	&	$-$0.051	&	-		    &	61.8(6.71)	&	-			&   - 	\\
5956.693    &       &   0.858   &   $-$4.505    &   -           &   -           &   48.4(4.98)  &   -           \\ 
6136.994	&		&	2.198	&	$-$2.95		&	-	    	&	51.0(6.84)	&	-			&   - 	\\
6137.694	&		&	2.588	&	$-$1.403	&	-	    	&	-		    &   108.5(4.87) &	40.4(4.65)			\\
6151.618    &       &   2.174   &   $-$3.370    &   -           &   -           &   20.5(5.18)  &   -           \\ 
6240.646	&		&	2.222	&	$-$3.380	&	-	    	&	38.6(6.78)	&	-			&   -	\\
6252.555    &       &   2.402   &   $-$1.690    &   -           &   -           &   133.7(5.16) &   -           \\ 
6254.258    &       &   2.277   &   $-$2.480    &   -           &   -           &   75.8(5.19)  &   -           \\ 
6297.800	&		&	2.222	&	$-$2.74		&	-	    	&	55.7(6.76)	&	-			&   -	\\
6318.018    &       &   2.452   &   $-$2.230    &   -           &   -           &   59.9(5.00)  &   -           \\ 
6335.328	&		&	2.198	&	$-$2.230	&	-		    &	-	    	&   79.7(4.81)  &	26.4(4.77)			\\
6408.016	&		&	3.687	&	$-$1.048	&	59.4(5.56)	&	-	    	&	-			&   -   	\\
6421.349	&		&	2.278	&	$-$2.027	&	-		    &	-		    &	91.4(4.88)  &   22.2(4.58)			\\
6430.85		&		&	2.18	&	$-$2.010	&	-		    &	-	    	&	-           &   26.5(4.52)			\\
4491.405	& Fe II	&	2.855	&	$-$2.700	&	57.6(5.68)	&	-		    &	-			&   -		\\
4508.288	&		&	2.855	&	$-$2.210	&	67.8(5.39)	&	-	    	&	-           &   29.2(4.63)			\\
4515.339	&		&	2.84	&	$-$2.480	&	62.5(5.53)	&	70.5(6.75)	&	22.9(5.11)  7   20.0(4.68)			\\
4520.224	&		&	2.81	&	$-$2.600	&	55.1(5.47)	&	-	    	&	-			&   -		\\
4629.339	&		&	2.807	&	$-$2.280	&	-		    &	77.8(6.74)	&	-			&   -		\\
4923.927    &       &   2.891   &   $-$1.260    &   -           &   -           &   80.7(4.93)  &   -           \\
5197.56		&		&	3.23	&	$-$2.250	&	-	    	&	72.2(6.74)	&	-			&   -		\\
\hline
\end{tabular}}

The numbers in the parenthesis in columns 5 - 8 give the derived absolute abundances from the respective line. 
  
\end{table*}
}

{\footnotesize
\begin{table*}
\caption{Equivalent widths (in m\r{A}) of lines used for deriving elemental abundances.} \label{linelist2}
\resizebox{\textwidth}{!}{\begin{tabular}{lccccccc}
\hline                       
Wavelength(\r{A}) & El & $E_{low}$(eV) & log gf & HE~0457$-$1805 & HE~0920$-$0506 & HE~1241$-$0337  &   HE~1327$-$2116   \\ 
\hline 
5682.633	& Na I	&	2.102	&	$-$0.700	&	140.1(6.49)	&	66.9(5.89)	&	-		    &   -       \\
5688.205	&		&	2.1	    &	$-$0.450	&	149.6(6.39)	&	-		    &   -           &	17.1(4.28)	\\
6154.226	&		&	2.102	&	$-$1.560	&	-	    	&	24.9(5.88)	&   42.1(5.48)  &	-		\\
6160.747	&		&	2.104	&	$-$1.260	&	112.3(6.51)	&	46.1(6.04)	&   45.2(5.23)  &	-		\\
4702.991	& Mg I	&	4.346	&	$-$0.666	&	144.4(6.85)	&	173.0(7.54)	&   -           &	-		\\
5528.405	&		&	4.346	&	$-$0.620	&	-		    &	165.6(7.47)	&   81.6(5.08)  &	65.0(5.22)	\\
5711.088	&		&	4.346	&	$-$1.833	&	106.2(7.04)	&	86.1(7.49)	&   16.5(5.26)  &	-		\\
5690.425	& Si I	&	4.929	&	$-$1.870	&	-		    &	32.5(7.02)	&   9.2(5.75)   &	-		\\
5772.148	&   	&	5.080	&	$-$1.750	&	-		    &	32.5(7.02)	&   10.7(5.87)  &	-		\\
5948.541	&		&	5.083	&	$-$1.23		&	-		    &	65.8(7.26)	&   -           &	-		\\
6145.016	&		&	5.616	&	$-$0.820	&	24.6(6.10)	&	-	    	&   -           &	-		\\
6237.319	&		&	5.613	&	$-$0.530	&	27.5(5.87)	&	-	    	&   -           &	10.2(5.45)	\\
4283.011	& Ca I	&	1.886	&	$-$0.224	&	114.3(5.09)	&	101.8(5.80)	&   -           &	-		\\
4318.652	&		&	1.899	&	$-$0.208	&	128.8(5.35)	&	-	    	&   -           &	-		\\
4425.437	&		&	1.879	&	$-$0.385	&	-	    	&	102.4(5.95)	&   -           &	-		\\
4435.679	&		&	1.89	&	$-$0.52		&	-	    	&	91.5(5.86)	&   -           &	-		\\
4456.616	&		&	1.899	&	$-$1.66		&	-		    &	58.7(6.12)	&   -           &	-		\\
4578.55		&		&	2.521	&	$-$0.56		&	73.9(5.22)	&	70.2(5.94)	&   -           &	17.2(4.17)	\\
5581.965	&		&	2.523	&	$-$1.833	&	-		    &	-	    	&   -           &	10.5(4.02)	\\
5590.114	&		&	2.521	&	$-$0.710	&	-		    &	-	    	&   28.4(4.06)  &	13.04.12)	\\
5594.462	&		&	2.523	&	$-$0.050	&	113.9(5.04)	&	-	    	&   -           &	-		\\
5857.451	&		&	2.932	&	0.23		&	-		    &	93.8(6.01)	&   -           &	36.8(4.17)	\\
6102.723	&		&	1.879	&	$-$0.890	&	-		    &	-		    &   -           &	44.2(4.18)	\\
6161.300    &       &   2.520   &   $-$1.270    &   -           &   -           &   17.4(4.31)  &   -               \\
6162.173	&		&	1.899	&	0.1		    &	-		    &	142.5(5.91)	&   165.5(4.31) &	-		\\
6166.439	&		&	2.521	&	$-$0.900	&	82.0(5.36)	&	57.9(5.87)	&   22.5(4.07)  &	-		\\
6169.042	&		&	2.523	&	$-$0.55		&	-		    &	78.2(6.04)	&   -           &	-		\\
6169.563	&		&	2.523	&	$-$0.27		&	-		    &	93.4(6.11)	&   -           &	-		\\
6439.075	&   	&	2.525	&	0.47		&	-		    &	136.8(6.03)	&   -           &	-		\\
6449.808	&		&	2.523	&	$-$0.550	&	98.6(5.25)	&	82.0(6.11)	&   -           &	-		\\
6455.598	&		&	2.523	&	$-$1.350	&	-		    &	47.9(6.05)	&   -           &	-		\\
6471.662	&		&	2.525	&	$-$0.590	&	-	    	&	80.6(6.12)	&   40.5(4.17)  &	-		\\
6493.781	&		&	2.521	&	0.14		&	134.9(5.16)	&	110.8(5.99)	&   -           &	-		\\
6499.65		&		&	2.523	&	$-$0.590	&	-		    &	68.7(5.81)	&   22.1(3.96)  &	-		\\
4453.312	& Ti I	&	1.43	&	$-$0.051	&	-	    	&	35.2(4.48)	&   -           &	-		\\
4512.734	&		&	0.836	&	$-$0.480	&	-		    &	55.7(4.44)	&   -           &	-		\\
4533.239	&		&	0.848	&	0.476		&	142.2(3.71)	&	-	    	&   -           &	-		\\
4617.269	&		&	1.749	&	0.389		&	-		    &	51.1(4.38)	&   -           &	12.9(2.76)	\\
4656.468	&		&	0	    &	$-$1.345	&	124.4(3.83)	&	-		    &   -           &	-		\\
4759.272	&		&	2.255	&	0.514		&	57.7(3.89)	&	33.8(4.30)	&   -           &	-		\\
4820.410    &       &   1.502   &   $-$0.441    &   -           &   -           &   21.3(2.79)  &   -       \\
4840.874	&		&	0.899	&	$-$0.509	&	-		    &	52.4(4.39)	&   -           &	-		\\
4913.622    &       &   1.870   &   0.166       &   -           &   -           &   31.7(2.88)  &   -       \\
5007.21		&		&	0.82	&	0.17		&	-		    &	-	    	&   131.5(2.65) &	68.5(2.79)	\\
\hline
\end{tabular}}

The numbers in the parenthesis in columns 5 - 8 give the derived absolute abundances from the respective line. \\
\end{table*}
}

{\footnotesize
\begin{table*}
\resizebox{\textwidth}{!}{\begin{tabular}{lccccccc}
\hline                       
Wavelength(\r{A}) & El & $E_{low}$(eV) & log gf & HE~0457$-$1805 & HE~0920$-$0506 & HE~1241$-$0337  & HE~1327$-$2116   \\ 
\hline 
5009.646	& Ti I	&	0.021	&	$-$2.259	&	81.4(3.92)	&	-	    	&   -           &	-		\\
5024.842	&   	&	0.818	&	$-$0.602	&	118.7(4.05)	&	56.9(4.51)	&   -           &	-		\\
5039.96		&		&	0.02	&	$-$1.13		&	-	    	&	-		    &   -           &	-		\\
5064.653	&		&	0.048	&	$-$0.991	&	139.7(3.56)	&	66.8(4.32)	&   -           &	-		\\
5210.386	&		&	0.047	&	$-$0.884	&	-	    	&	71.2(4.37)	&   -           &	-		\\
4161.535	& Ti II	&	1.084	&	$-$2.360	&	-		&	-		        &   -           &	41.9(2.51)	\\
4417.719	&		&	1.165	&	$-$1.430	&	-		&	84.4(4.27)	    &   -           &	-		\\
4418.33		&		&	1.237	&	$-$2.460	&	-		&	68.9(4.41)	    &   -           &	-		\\
4443.794	&		&	1.08	&	$-$0.700	&	-		&	130.6(4.40)	    &   -           &	-		\\
4468.52		&		&	1.13	&	$-$0.6		&	-		&	134.9(4.41)	    &   -           &	-		\\
4493.51		&		&	1.08	&	$-$2.73		&	-		&	-		        &   -           &	17.8(2.58)	\\
4568.314	&		&	1.22	&	$-$2.650	&	59.4(3.30)	&	-		    &   -           &	-		\\
4571.96		&		&	1.571	&	$-$0.53		&	135.1(3.25)	&	-		    &   -           &	-		\\
4764.526	&		&	1.236	&	$-$2.770	&	64.9(3.48)	&	40.0(4.26)	&   -           &	-		\\
4779.985	&		&	2.048	&	$-$1.37		&	-		&	-		        &   -           &	37.0(2.70)	\\
4798.521	&		&	1.08	&	$-$2.43		&	-		&	64.9(4.44)	    &   -           &	-		\\
4865.61 	&		&	1.12	&	$-$2.70		&	-		&	-       	    &   32.6(2.62)  &	-		\\
5185.9		&		&	1.89	&	$-$1.35		&	94.4(3.37)	&	65.8(4.23)	&   63.0(2.75)  &	-		\\
5226.543	&		&	1.566	&	$-$1.300	&	-		&	81.0(4.30)	    &   77.7(2.39)  &	-		\\
5336.77     &       &   1.58    &   $-$1.700    &   -       &   -               &   43.8(2.34)  &   -       \\
5381.015	&		&	1.566	&	$-$2.08		&	94.1(3.61)	&	-		    &   -           &	23.5(2.55)	\\
4351.05		& Cr I	&	0.97	&	$-$1.45		&	-		&	67.0(5.21)	    &   -           &	-		\\
4616.12		&		&	0.982	&	$-$1.190	&	-		&	80.6(5.33)	    &   -           &	-		\\
4652.157	&		&	1.004	&	$-$1.030	&	-		&	82.5(5.24)	    &   -           &	-		\\
4737.347	&		&	3.087	&	$-$0.099	&	-		&	50.9(5.46)	    &   -           &	-		\\
4829.372	&		&	2.544	&	$-$0.810	&	-		&	46.4(5.47)	    &   -           &	-		\\
5247.565	&		&	0.961	&	$-$1.640	&	134.1(4.71)	&	68.8(5.28)	&   -           &	-		\\
5296.691	&		&	0.982	&	$-$1.400	&	-		&	-		        &   -           &	21.6(3.03)	\\
5298.277	&		&	0.983	&	$-$1.15		&	-		&	-		        &   -           &	34.1(3.02)	\\
5300.744	&		&	0.982	&	$-$2.120	&	104.1(4.70)	&	-	    	&   -           &	-		\\
5312.871	&		&	3.45	&	$-$0.562	&	21.8(4.87)	&	-		    &   -           &	-		\\
5345.801	&		&	1.003	&	$-$0.980	&	-		&	94.2(5.37)	    &   -           &	-		\\
5348.312	&		&	1.003	&	$-$1.290	&	148.6(4.62)	&	84.7(5.43)	&   -           &	-		\\
5409.772	&		&	1.03	&	$-$0.720	&	-		&	107.8(5.42) 	&   -           &	-		\\
5787.965	&		&	3.323	&	$-$0.083	&	-		&	41.4(5.37)	    &   -           &	-		\\
4588.19		& Cr II	&	4.072	&	$-$0.63		&	81.1(4.55)	&	-		    &   24.4(3.57)  &	-		\\
4634.07		&		&	4.072	&	$-$1.240	&	-		&	58.0(5.41)	    &   -           &	-		\\
4848.25		&		&	3.864	&	$-$1.140	&	46.5(4.60)	&	-		    &   23.6(3.53)  &	-		\\
5305.853	&		&	3.827	&	$-$2.357	&	-		&	30.9(5.42)	    &   -           &	-		\\
5334.869	&		&	4.073	&	$-$1.562	&	-		&	45.4(5.29)	    &   -           &	-		\\
4686.207	& Ni I	&	3.597	&	$-$0.64		&	62.8(5.19)	&	-		    &   -           &	-		\\
4703.803	&		&	3.658	&	$-$0.735	&	-		&	45.1(5.87)	    &   -           &	-		\\
4731.793	&		&	3.833	&	$-$0.85		&	-		&	32.9(5.80)	    &   -           &	-		\\
4732.46		&		&	4.106	&	$-$0.55		&	-		&	-		        &   -           &	10.1(4.65)	\\
\hline
\end{tabular}}

The numbers in the parenthesis in columns 5 - 8 give the derived absolute abundances from the respective line. \\
\end{table*}
}

{\footnotesize
\begin{table*}
\resizebox{\textwidth}{!}{\begin{tabular}{lccccccc}
\hline                       
Wavelength(\r{A}) & El & $E_{low}$(eV) & log gf & HE~0457$-$1805 & HE~0920$-$0506 & HE~1241$-$0337 & HE~1327$-$2116   \\ 
\hline 
4752.415	& Ni I	&	3.658	&	$-$0.7		&	-		&	46.8(5.88)	    &   -           &	-		\\
4756.51		&   	&	3.48	&	$-$0.340	&	94.1(5.39)	&	66.3(5.96)	&   -           &	-		\\
4821.13		&		&	4.153	&	$-$0.85		&	23.3(5.25)	&	-		    &   -           &	-		\\
4852.56		&		&	3.542	&	$-$1.07		&	-		&	38.0(5.86)	    &   -           &	16.7(4.76)	\\
4855.41     &       &   3.54    &   0.000       &   -           &   -           &   27.9(3.51)  &   -           \\
4937.34     &       &   3.60    &   $-$0.390    &   -           &   -           &   20.5(3.80)  &   -           \\
4953.2		&		&	3.74	&	$-$0.67		&	66.5(5.41)	&	42.8(5.81)	&   -           &	-		\\
5035.357	&		&	3.635	&	0.29		&	-		&	82.2(5.89)	    &   -           &	-		\\
5081.11     &       &   3.84    &   0.300       &   -           &   -           &   25.3(3.52)  &   -           \\
5146.48		&		&	3.706	&	0.12		&	-		&	64.9(5.46)	    &   -           &	-		\\
6176.807	&		&	4.088	&	$-$0.260	&	-		&	-		        &   -           &	13.8(4.67)	\\
6177.236	&		&	1.826	&	$-$3.5		&	57.7(5.38)	&	-		    &   -           &	-		\\
6186.71		&		&	4.106	&	$-$0.777	&	37.9(5.28)	&	-		    &   -           &	-		\\
6327.593	&		&	1.676	&	$-$3.150	&	-		&	29.4(5.65)	    &   -           &	-		\\
6378.247	&		&	4.154	&	$-$0.89		&	29.3(5.26)	&	24.6(5.88)	&   -           &	-		\\
6643.63     &       &   1.68    &   $-$2.220    &   -           &   -           &   84.7(3.74)  &   -           \\
4722.15		& Zn I	&	4.029	&	$-$0.370	&	-		&	68.2(4.25)	    &   -           &	-		\\
4810.53		&		&	4.08	&	$-$0.170	&	69.9(3.10)	&	-		    &   12.4(1.81)  &	22.5(2.16)	\\
6362.34		&		&	5.8	&	0.15		&	-		&	29.2(4.42)	        &   -           &	-		\\
4883.684	& Y II	&	1.084	&	0.07		&	-		&	103.9(2.81)	    &   134(0.73)   &	79.7(0.08)	\\
5087.416	&		&	1.084	&	$-$0.170	&	-		&	90.8(2.76)	    &   126.8(0.78) &	59.1(0.06)	\\
5119.11     &       &   0.99    &   $-$1.360    &   -       &   -               &   64.8(0.97)  &   -           \\
5200.406	&		&	0.992	&	$-$1.360	&	-		&	78.0(2.71)	    &   -           &	-		\\
5205.724	&		&	1.033	&	$-$0.34		&	-		&	79.1(2.55)	    &   -           &	-		\\
5402.774	&		&	1.839	&	$-$0.51		&	-		&	49.8(2.56)	    &   -           &	-		\\
5544.611	&		&	1.738	&	$-$1.090	&	-		&	38.3(2.62)	    &   -           &	-		\\
5546.01     &       &   1.75    &   $-$1.100    &   -       &   -               &   21.8(0.98)  &   -       \\
5662.925	&		&	1.944	&	0.16		&	-		&	-		        &   79.0(0.83)  &	17.7(0.04)	\\
4317.31     & Zr II &   0.71    &   $-$1.450    &   -       &   -               &   77.9(1.55)  &   -           \\
4379.74     &       &   1.53    &   $-$0.360    &   -       &   -               &   79.9(1.59)  &   -           \\
5112.27     &       &   1.67    &   $-$0.85     &   -       &   -               &   42.4(1.51)  &   -           \\
5350.09     &       &   1.82    &   $-$1.24     &   -       &   -               &   14.6(1.50)  &   -           \\
4193.87     & Ce II &   0.55    &   $-$0.480    &   -       &   -               &   41.3(0.36)  &   -           \\
4336.244	&   	&	0.704	&	$-$0.564	&	-		&	-		        &   -           &	20.1(0.52)	\\
4349.789	&		&	0.701	&	$-$0.107	&	-		&	44.3(2.11)	    &   -           &	-		\\
4364.653	&		&	0.495	&	$-$0.201	&	-		&	41.9(1.86)	    &   -           &	-		\\
4418.78		&		&	0.863	&	0.177		&	-		&	41.1(1.88)	    &   -           &	-		\\
4427.916	&		&	0.535	&	$-$0.460	&	107.5(1.83)	&	-		    &    32.7(0.05) &	32.8(0.46)	\\
4460.207	&		&	0.477	&	0.171		&	-		&	-		        &   -           &	98.0(0.54)	\\
4483.893	&		&	0.864	&	0.01		&	124.4(2.16)	&	-		    &   -           &	-		\\
4486.909	&		&	0.295	&	$-$0.474	&	-		&	48.3(2.19)	    &   -           &	-		\\
4508.079	&		&	0.621	&	$-$1.238	&	70.8(1.84)	&	-		    &   -           &	-		\\
4560.28		&		&	0.91	&	0		&	-		&	-		            &   -           &	43.1(0.58)	\\
4628.161	&		&	0.516	&	0.008		&	-		&	-		        &   -           &	62.2(0.33)	\\
4873.999	&		&	1.107	&	$-$0.892	&	61.7(1.83)	&	-		    &   -           &	-		\\
5187.458	&		&	1.211	&	$-$0.104	&	102.7(1.94)	&	-		    &   -           &	-		\\
5274.23     &       &   1.04    &   0.130       &   -           &   -           &   56.2(0.39)  &   -           \\
6034.205	&		&	1.458	&	$-$1.019	&	55.5(2.10)	&	-		    &   -           &	-		\\
5188.217	& Pr II	&	0.922	&	$-$1.145	&	41.5(1.25)	&	-		    &   -           &	-		\\
5219.045	&		&	0.795	&	$-$0.24		&	101.5(1.31)	&	-		    &   20.7($-$0.32) &	-		\\
5259.728	&		&	0.633	&	$-$0.682	&	91.1(1.31)	&	12.1(1.12)	&   -           &	-		\\
5292.619	&		&	0.648	&	$-$0.300	&	104.6(1.21)	&	22.8(1.17)	&   -           &	-		\\
5322.772	&		&	0.482	&	$-$0.315	&	131.8(1.51)	&	-		    &   34.3($-$0.41)  &	20.4($-$0.23)	\\
6165.891	&		&	0.923	&	$-$0.205	&	102.4(1.22)	&	-		    &   12.0($-$0.55) &	-		\\
\hline
\end{tabular}}

The numbers in the parenthesis in columns 5 - 8 give the derived absolute abundances from the respective line. 
\end{table*}
}

{\footnotesize
\begin{table*}
\resizebox{\textwidth}{!}{\begin{tabular}{lccccccc}
\hline                       
Wavelength(\r{A}) & El & $E_{low}$(eV) & log gf & HE~0457$-$1805 & HE~0920$-$0506 & HE~1241$-$0337  & HE~1327$-$2116   \\ 
\hline 
4446.384	& Nd II	&	0.204	&	$-$0.590	&	139.7(1.99)	&	-		    &   -           &	-		\\
4451.563	&		&	0.38	&	$-$0.040	&	145.4(1.76)	&	-		    &   -           &	-		\\
4811.342	&		&	0.064	&	$-$1.140	&	127.3(1.82)	&	38.4(1.91)	&   49.8(0.08)  &	47.7(0.52)	\\
4825.478	&		&	0.182	&	$-$0.86		&	-		&	34.9(1.62)	    &   -           &	44.0(0.33)	\\
4859.039	&		&	0.32	&	$-$0.83		&	-		&	-	        	&   57.8($-$0.17) &	38.8(0.39)	\\
5130.59		&		&	1.3	&	0.1		&	-		&	33.4(1.79)	            &   -           &	-		\\
5212.361	&		&	0.204	&	$-$0.870	&	137.4(1.76)	&	-		    &   -           &	-		\\
5255.506	&		&	0.204	&	$-$0.820	&	-		&	-		        &   -           &	51.2(0.36)	\\
5287.133	&		&	0.744	&	$-$1.300	&	78.1(1.74)	&	-		    &   -           &	-		\\
5293.163	&		&	0.822	&	$-$0.060	&	-		&	-		        &   59.6($-$0.08)   &	51.2(0.34)	\\
5311.453	&   	&	0.986	&	$-$0.420	&	93.8(1.63)	&	-		    &   -           &	-		\\
5319.815	&		&	0.55	&	$-$0.210	&	-		&	43.2(1.64)	    &   -           &	59.3(0.25)	\\
5361.51		&		&	0.68	&	$-$0.4		&	-		&	-		        &   -           &	36.0(0.30)	\\
5442.264	&		&	0.68	&	$-$0.910	&	115.8(1.95)	&	-		    &   -           &	-		\\
5485.696	&		&	1.264	&	$-$0.120	&	107.2(1.79)	&	-		    &   -           &	-		\\
5688.518	&		&	0.986	&	$-$0.25		&	-		&	-		        &   29.8(0.06)  &	32.8(0.43)	\\
5825.857	&		&	1.08	&	$-$0.760	&	95.2(1.85)	&	-		    &   -           &	-		\\
4318.927	& Sm II	&	0.28	&	$-$0.270	&	-		&	-		        &   55.6($-$0.51) &	55.1($-$0.19)	\\
4424.337	&		&	0.485	&	$-$0.26		&	-		&	47.0(1.68)	    &   65.5($-$0.53) &	-		\\
4434.318	&		&	0.378	&	$-$0.576	&	-		&	-		        &   -               &	24.8($-$0.24)	\\
4499.475	&		&	0.248	&	$-$1.413	&	81.5(1.29)	&	-		    &   -               &	-		\\
4566.21		&		&	0.33	&	$-$1.245	&	88.7(1.36)	&	22.5(1.46)	&   -               &	-		\\
4577.69		&		&	0.25	&	$-$0.77		&	-		&	-		        &   -               &	33.0($-$0.07)	\\
4642.228	&		&	0.379	&	$-$0.951	&	-		&	27.4(1.39)	    &   -               &	23.6(0.08)	\\
4674.593	&		&	0.184	&	$-$1.055	&	116.9(1.54)	&	-		    &   -               &	-		\\
4676.902	&		&	0.04	&	$-$1.407	&	-		&	-		        &   -               &	21.6(0.08)	\\
4704.4		&		&	0	&	$-$1.562	&	102.5(1.46)	&	32.0(1.75)	    &   -               &	-		\\
4791.58		&		&	0.104	&	$-$1.846	&	81.6(1.41)	&	-		    &   -               &	-		\\
4844.209	&		&	0.277	&	$-$1.558	&	87.3(1.46)	&	-		    &   -               &	-		\\
4854.368	&		&	0.379	&	$-$1.873	&	53.0(1.23)	&	-		    &   -               &	-		\\
\hline
\end{tabular}}

The numbers in the parenthesis in columns 5 - 8 give the derived absolute abundances from the respective line. 
\end{table*}
}

{\footnotesize
\begin{table*}
\caption{Lines used to derive the elemental abundance from spectral synthesis.} \label{linelist3}
\resizebox{\textwidth}{!}{\begin{tabular}{lccccccc}
\hline                       
Wavelength(\r{A}) & El & $E_{low}$(eV) & log gf & HE~0457$-$1805 & HE~0920$-$0506 & HE~1241$-$0337  &   HE~1327$-$2116   \\ 
\hline 
4320.730    & Sc II &   0.600   &   $-$1.950    &   -           &   -           &  0.91         &   -      \\
4374.457    &       &   0.620   &   $-$0.440    &   -           &   -           &   -           & 0.04      \\
4415.560    &       &   0.590   &   $-$1.700    &   -           &   -           &   1.09        &   -      \\
5031.021	&       &	1.357	&   $-$0.260	&   -		    &   -           &   -           & 0.02		\\
6245.637	&       &   1.507	&   $-$0.980	& 1.85	        & 2.14          &   -		    & 	-		\\
6604.600    &       &   1.357   &   $-$1.480    &   -           & 2.14          &   -           &   -      \\
4864.731    & V I   &   0.017   &   $-$0.960    & 2.73          &   -           &   -           &   -      \\
5727.048    &       &   1.080   &   $-$0.914    & 2.73          & 2.87          &   -           &   -       \\
4451.586    & Mn I  &   2.889   &   0.278       &   -           &   -           &   -           & 3.63      \\
4470.140    &       &   2.941   &   $-$0.444    &   -           &   -           &   -           & 3.59      \\
5516.774    &       &   2.178   &   $-$1.847    &   -           &   -           & 3.55          &   -       \\
6013.513    &       &   3.072   &   $-$0.251    & 4.43          & 4.43          &   -           &   -       \\
6021.819    &       &   3.075   &   0.0340      & 4.43          & 4.45          &   -           &   -       \\
4118.770    & Co I  &   1.050   &   $-$0.490    &   -           &   -           &   -           & 3.59      \\
4121.320    &       &   0.920   &   $-$0.320    &   -           &   -           &   -           & 3.63      \\
5342695     &       &   4.021   &   0.690       & 3.58          &   -           &   -           &   -       \\
5483.344    &       &   1.711   &   $-$1.490    &   -           & 4.19          &   -           &   -       \\
5105.537    & Cu I  &   1.389   &   $-$1.516    & 2.21          & 3.19          &   -           &   -       \\
7800.259    & Rb I  &   0.000   &   0.140       & 1.95          & 1.70          &   -           &   -       \\
4607.327    & Sr I  &   0.000   &   0.283       & 2.54          & 3.57          & 1.54          &   -       \\ 
6435.004    & Y I   &   0.066   &   $-$0.820    & 3.01          & 2.32          &   -           &   -       \\
6134.585    & Zr I  &   0.000   &   $-$1.280    & 3.03          & 3.05          & 1.44          & 1.81      \\
5853.668    & Ba II &   0.295   &   $-$0.840    & 2.73          & 2.83          & 0.75          & 1.08      \\
6496.897    &       &   0.604   &   $-$1.886    &   -           & 2.83          &   -           &   -       \\
4748.726    & La II &   0.926   &   $-$0.540    & 1.37          & 1.60          &   -           &   -       \\
4921.776    &       &   0.244   &   $-$0.450    & 1.40          & 1.60          & $-$0.35       &   -       \\
5259.379    &       &   0.173   &   $-$1.950    & 1.40          &   -           &   -           &   -       \\
5303.528    &       &   0.321   &   $-$2.246    &   -           &   -           &   -           & 0.00      \\
4129.725    & Eu II &   0.000   &   $-$1.294    &   -           & $-$0.43       &   -           &   -       \\
6437.640    &       &   1.319   &   $-$1.998    &   -           &   -           & $-$1.58       &   -       \\
6645.064    &       &   1.379   &   $-$0.517    & $-$0.20       & $-$0.39       & $-$1.28       & $-$1.16   \\
\hline
\end{tabular}}

The numbers in columns 5 - 8 give the derived absolute abundances from the respective line. \\
\end{table*}
}

\end{document}